\documentclass[12pt, draftclsnofoot, onecolumn]{IEEEtran}
\usepackage{stfloats}
\ifCLASSINFOpdf
\else
\fi

\usepackage{CJK}
\usepackage{graphicx}
\usepackage{caption}
\usepackage{epstopdf}
\usepackage{subfigure}
\usepackage[cmex10]{amsmath}
\usepackage[amsmath,thmmarks]{ntheorem}
\usepackage{amssymb}
\usepackage{ntheorem}
\usepackage{mathabx}
\usepackage{algorithm}
\usepackage{algorithmic}
\usepackage{multirow}
\usepackage{xcolor}
\usepackage{multirow}
\usepackage{bm}
\usepackage{mathabx}
\usepackage{stmaryrd}
\usepackage{upgreek}
\begin{document}
\title{Energy-delay-cost Tradeoff for Task Offloading in Imbalanced Edge Cloud Based Computing}

\author{Weiheng~Jiang,~
        Yi~Gong,~Yang~Cao,~Xiaogang~Wu~and~Qian~Xiao
\thanks{W. Jiang, X. Wu and Q. Xiao are with the Center of Communication and Tracking Telemetering Command/College of Communication Engineering, Chongqing University, Chongqing 400044. W. Jiang is also with the Department of Electrical and Electronic Engineering, Southern University of Science and Technology, Shenzhen 518055, China (e-mail: whjiang@cqu.edu.cn, xiaogangwu@cqu.edu.cn, qxiao@cqu.edu.cn).}
\thanks{Y. Gong is with the Department of Electrical and Electronic Engineering, Southern University of Science and Technology, Shenzhen 518055, China (e-mail: gongy@sustc.edu.cn).}
\thanks{Y. Cao is with the Wuhan National Laboratory for Optoelectronics, School of Electronics Information and Communications, Huazhong University of Science and Technology, Wuhan 430074, China (e-mail: ycao@hust.edu.cn).}}

\maketitle

\begin{abstract}
In this paper, the imbalance edge cloud based computing offloading for multiple mobile users (MUs) with multiple tasks per MU is studied. In which, several edge cloud servers (ECSs) are shared and accessed by multiple wireless access points (APs) with the backhaul links, and each MU has multiple computing intensive and latency critical tasks should be offloaded to execute, which involves both the AP association and ECS selection, with the objective of minimizing the offloading cost. Distinguished with existing research, besides the transmission delay and energy consumption from MU to AP, the ECS access-cost which characterizes the ECS access delay and (or) resource using cost is introduced, thus finally formulates the delay-energy-cost tradeoff based offloading cost criteria, for the MUs' offloading decision and resource allocation problems. In specific, in our system, the ECS access-cost depends on both the AP and ECS, which reflects the effects of different backhaul techniques used by these APs, and their negotiated payments of ECS resource using under different service level agreements. Both problems of minimizing the sum offloading costs for all MUs (efficiency-based) and minimizing the maximal offloading cost per MU (fairness-based) are discussed. Since these problems are all NP-hard. Therefore, some centralized and distributed heuristic algorithms are proposed to find the suboptimal solutions. Further analysis and numerical results are presented at last to demonstrate the performance of these algorithms from several aspects, such as efficiency, complexity and fairness.
\end{abstract}


\IEEEpeerreviewmaketitle

\section{Introduction}
\subsection{Background and the state-of-art}
\IEEEPARstart{O}{ver} the past decade or so, we have witnessed the technological innovation of mobile communication and networks. Recently, the industrial organizations, academic institutes and standard committees are focused on the critical technologies of the fifth generation mobile communication system (5G). However, in order to motivate more innovations in the field of mobile internet, information and communication technology (ICT) community has to tackle the challenges from mobile devices, massive internet-of-things (IoTs) applications and the operators.

From the perspective of mobile devices, the demands of latency critical and computing intensive applications are increased i.e., AR/VR, multimedia IoT based cooperative video processing \cite{42} et al. These applications are hoped to be executed with both lower latency and less energy consumption over the mobile devices, to win better user experience. However, due to the constraints of cost, size and weight at the mobile device, its computing capability and energy supply are restricted \cite{1}, and the technique breakthrough seems unpredictable in the near future \cite{2}. From the perspective of IoTs, the development of Internet-of-Everything (IoE) brings massive IoT based cloud connection requests. Though the remote cloud has abundant computing and storage resources, its access capability is poor. It is predicted by Cisco that at the end of 2020, about 50 billion IoT devices (sensors, or wearable devices) need to interact with the cloud \cite{3}. On the one hand, the large number of cloud connections will cause the congestion of backhual, on the other hand, the remote cloud service requests from these devices have to cross the wide area network (WAN), thus it is difficult to promise the delay and jitter based quality-of-service (QoS). From the perspective of operators, though their investments and the amount of network traffic are increasing, their developed networks are gradually channelized and the average revenue per user (ARPU) is constantly decrease \cite{4}.

In order to overcome above contradictions, a new technology named mobile edge computing (MEC) is proposed \cite{5}, \cite{6} and it is known to be one of the key tools for the coming 5G. In essence, MEC is a novel paradigm who extends the centralized cloud computing capabilities to the edge of cloud. OpenFog Consortium and standard development organizations like ETSI have also recognized the benefits that the MEC can bring to consumers \cite{7}. In particular, it is noted that MEC is becoming an important enabler of consumer-centric applications and services that demand real-time operations, e.g., smart mobility, connected vehicles, smart cities, and location based services. For our prementioned challenges, leveraging the MEC, mobile devices or mobile users (MUs) can offload their computing intensive and latency critical tasks to the edge cloud server (ECS), thus significantly lower the computing capability requirements at the mobile device and reduce its energy consumption caused by local task implementation. In addition, with MEC, massive IoT devices do not need to access remote cloud but the nearby ECS thus relieve the stress of backhual link. Last but not the least, by introducing the ECS in the system, the mobile network operators (MNOs) can deploy value-added service over it and open up the platform to the third parties.

In the MEC based system, computing offloading renders the communication and computing are coupled with each other, and the user-perceived performance is determined by both the offloading decision and the jointly computing and communication resource allocation. On the one hand, the offloading decision and resource allocation depends on the properties of the task, i.e., binary offloadable task \cite{8}-\cite{13}, or partial offloadable task \cite{14}-\cite{16}, \cite{21}, \cite{25}. For the former, the tasks are either implemented locally or offloaded to the cloud to execute; while for the latter, a task can be splitted into at least two parts, and one of them is performed at local device and the other is offloaded to the cloud platform. On the other hand, offloading decision and resource allocation are affected by the system access scheme, who determines how MUs sharing both the communication and computing resource in the system. For the communication resource, both orthogonal access scheme, i.e., time-division multiple access (TDMA) \cite{14} and orthogonal frequency division multiple access (OFDMA) \cite{11}, \cite{14}, \cite{17}, \cite{18}, and non-orthogonal access scheme, i.e., code-division multiple access (CDMA) \cite{19}, \cite{20} and non-orthogonal multiple access (NOMA) \cite{21} have been discussed. For the computing resource, depending on how many virtual machines cloud be virtualized by the system, multiple tasks could be serially or concurrently executed at the cloud. In addition, the system design objective or the user-perceived performance also plays a leading role in the offloading decision and resource allocation. Typically system design objectives including latency minimization \cite{15}, \cite{24}, energy consumption reducing \cite{14}, \cite{15}, \cite{17}, \cite{18}, \cite{21}, \cite{22}, \cite{25}, and delay-energy tradeoff \cite{8}-\cite{13}, \cite{19}, \cite{20}, \cite{23}. A comprehensive survey of the offloading decision and resource allocation for MEC system by considering other system scenarios and configurations can be found in \cite{4}, \cite{26} and \cite{27}.

\subsection{Motivations and Contributions}
The motivations of our work come from following three aspects.

At first, in discussing the offloading decision and resource allocation problem, existing works assumed that the edged cloud servers (ECSs) of the MEC system are balanced deployment, i.e., each wireless access point (AP) or base station (BS) is equipped with an independent ECS \cite{12}-\cite{15}, or multiple APs access into the unique ECS \cite{12}-\cite{17}. However, in practical, by considering the inhomogeneous traffic distribution over space and the expensive deployment cost for the ECS, the imbalanced ECS deployment strategy is more attractive for the operators, as Fig. 1(a), i.e., multiple APs share and access into several ECSs by multi/single-hop backhaul links \cite{26}. Till now, few attention has been paid for this scenario, and in which the resulted problem of offloading decision and resource allocation is complicated by the coupled relationship between shared ECSs and multiple APs.

Secondly, in most literatures, latency \cite{12}, energy \cite{10}, \cite{14} or delay-energy tradeoff \cite{8}, \cite{9}, \cite{12}, \cite{13} is chosen as the system performance criteria but ignore the cost for accessing the ECS. In our imbalanced MEC system, an additional ECS access-cost is introduced to characterize the factors such as ECS access delay \cite{28}, or ECS using payments by considering possible different service level agreements (SLAs) between the mobile virtual operators (MVOs) and the cloud service providers\footnote{Our proposed schemes are particularly suitable for the scenario of virtualized wireless network with multi-tenants, in which the mobile virtualized network operators (MVNOs) lease cloud resource from the cloud service providers \cite{30}.} \cite{29}, \cite{30}. In specific, this cost is jointly determined by the accessed AP and the ECS. In this case, the offloading decisions of MUs should jointly consider the delay-energy-cost tradeoff to make the best decision. This issue can be explained by an example shown in Fig. 1(b), while the system has four MUs (we assume one task per MU for simplify), i.e., $s_1$, $s_2$, $s_3$ and $s_4$, and three APs, i.e., $b_1$, $b_2$ and $b_3$, and two ECS, i.e., $c_1$ and $c_2$. For each MU $s_i, i=1, ..., 4$, its task is characterized by a quadruple $(x_1^i,x_2^i,x_3^i,x_4^i)$. In which, $x_1^i$ denotes the CPU resource requirement of the task, $x_2^i$, $x_3^i$ and $x_4^i$ represent the weighted sum of delay and energy for $s_i$'s task is offloaded by AP $b_1$, $b_2$, and $b_3$, respectively. For instance, $s_1(3,2,2,5)$ denotes the unit of CPU resource required for $s_1$'s task is 3, and the weighted sum of delay and energy for the task be offloaded by $b_1$, $b_2$, and $b_3$ are 3, 2 and 5, respectively. The AP $b_j,j=1,2,3$ is characterized by a tuple $(y_1^j,y_2^j)$, which denotes the ECS access-cost of $b_j$ to $c_1$ and $c_2$, e.g., $b_1(2,3)$ means that the ECS access-cost of $b_1$ to $c_1$ and $c_2$ are 2 and 5, respectively. For the ECS $c_k, k=1,2$, which is characterized by $(z^k)$, denoting the available CPU resource at $c_k$, e.g., $c_1(5)$ represents that the $c_1$ has 5 units of CPU resource. One can note that, different offloading decision (or path selection) have varying user-perceived performance, i.e., for the offloading path $s_2-b_1-c_2$, $s_2$ offloads its task by $b_1$ to $c_2$, the sum of delay, energy and cost is 4. While for the offloading path $s_2-b_3-c_2$, $s_2$ offloads its task by $b_3$ to $c_2$, the sum of delay, energy and cost is 5. The offloading decisions of MUs are affected by many factors, such as the offloading delay and energy consumption for an AP, the ECS access-cost, the available CPU resources at the ECS and the strategies of other MUs.

\begin{figure}[]
\centering
\includegraphics[width=3.5in]{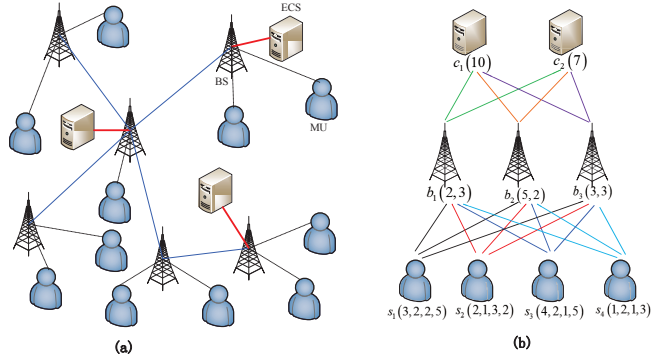}
\caption{Imbalanced edge cloud system example.}
\label{fig.1}
\end{figure}

The last motivation of this work is about the system design criteria. For the scenario of multi-MUs with multi-tasks per MU, existing works still focus on minimizing the weighted sum of delay, energy or delay-energy tradeoff. It is no doubt that the resulted solution may cause unfairness between MUs, i.e., some MUs always have good offloading performance for all their tasks while other MUs always have bad offloading performance. To improve the fairness of the offloading decision and resource allocation, fairness based criteria should be introduced in the system design. In summary, the contributions of this paper are as follows.

First, by considering the imbalanced ECS deployment scenario, the jointly offloading decision and resource allocation for multi-MUs with multi-tasks per MU is studied, in which multiple APs are not directly connected with an independent ECS but shared and accessed into several ECSs by multi/single-hop backhaul links. The considered scenario is more practical compared with the existing balanced ECS deployment assumptions.

Second, the delay-energy-cost tradeoff based offloading cost is proposed in our work to quantify the user-perceived performance of the task offloading, and then it is used to conduct the jointly task offloading decision and resource allocation. Moreover, the ECS access-cost depends on both the associated AP and accessed ECS, and which reflects the different ECS access delay and service (or resource using) cost negotiated by APs and ECSs.

At last, both problems of minimizing the sum of offloading cost for all MUs (efficiency-based) and minimizing the maximal offloading cost per MU (fairness-based) are formulated and discussed, i.e., OP1 and OP2. Since these problems are all NP-hard, two linear relaxation algorithms are proposed as the benchmark schemes (or as the performance bound), i.e., linear relaxation algorithm for efficiency problem (ELR) and linear relaxation algorithm for fairness problem (FLR). Then for OP1, with different algorithm implementation limitations, the centralized greed algorithm (CGA), the modified greedy algorithm (MGA) and asynchronous distributed matching algorithm (ADMA) are proposed to obtain its suboptimal solution. For OP2, based on CGA, a fairness based greedy algorithm (FGA) is proposed. The relationship among these algorithms is summarized in Fig. 2. Numerical simulations have been done at last to demonstrate the performance of these algorithms.

\begin{figure}[]
\centering
\includegraphics[width=3.5in]{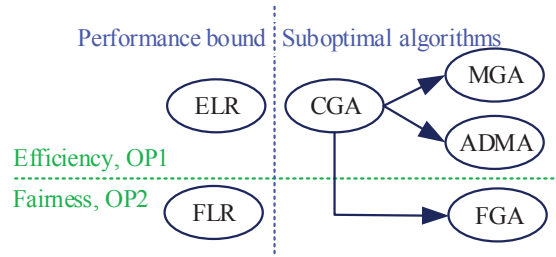}
\caption{The work in this paper.}
\label{fig.2}
\end{figure}
The organization of the remainder paper is as follows. In Section II, we present the system model and formulate the discussed problems. In section IV, the efficiency-based problem, i.e., minimizing the sum of offloading cost for all MUs, is discussed and solved, both the centralized and distributed suboptimal algorithms are proposed therein. The fairness based problem is further discussed and tackled in Section V. Some numerical results of the proposed algorithms are given in Section VI and we conclude at last.

\section{System model and problems}
In this section, we first present the system model. Then, the discussed problems are formulated.

\subsection{System model}
In this paper, an imbalance edge cloud system which including $|\mathcal{A}|$ mobile users (MUs), $|\mathcal{B}|$ wireless access points (APs) and $|\mathcal{C}|$ edge-based cloud servers (ECSs) is discussed, as shown in the Fig. 1. $\mathcal{A}=\{a_1,a_2,...,a_{|\mathcal{A}|}\}$, $\mathcal{B}=\{b_1,b_2,...,b_{|\mathcal{B}|}\}$ and $\mathcal{C}=\{c_1,c_2,...,c_{|\mathcal{C}|}\}$ denote the sets of MUs, APs and ECSs, respectively\footnote{In this paper, we will alternatively use $i$ or $a_i$ to denote the $i$th MU, $j$ or $s_{i,j}$ to denote the $j$th task of MU $i$, $m$ or $b_m$ to denote the $m$th AP, and $n$ or $c_n$ to denote the $n$th ECS.}. For each MU, there are several offloadable tasks to be executed. We use $\mathcal{S}_i=\{s_{i,1},s_{i,2},...,s_{i,S_i}\}$ and $r_{i,j}$ to denote the task set of MU $a_i$ and the computing resource requirement for its task $s_{i,j}, i= 1,...,|\mathcal{A}|, j =1, ..., S_i$, respectively. As \cite{11} and \cite{17}, these MUs are resource-hungry devices and have limited computing capability or energy supply. Therefore, all these computing-intensive tasks must be offloaded by these APs to ECS to execute. Our work can be directly extended to the scenario that the MUs could locally execute the tasks.

As mentioned earlier, by considering more practical scenario, in the imbalanced edge cloud system, each AP is not directly equipped with an ECS, but connected to the shared ECS through backhual links\footnote{In practical, we have $|\mathcal{C}|<|\mathcal{B}|$, i.e., the available ECS is less than APs for cost-efficiency based ECS deployment \cite{28}. In addition, though mostly we assume that each AP accesses into all ECSs in the system, our proposed algorithms are still valid for the scenario that each AP is connected with only a subset of the ECSs.}. However, these APs may incur different ECS access-costs due to the access-cost comes from the ECS access delay or (and) resource using payment. We use $\delta_{m,n}$ to denote the access-cost from AP $b_m (m=1,...,|\mathcal{B}|)$ to ECS $c_n (n=1,...,|\mathcal{C}|)$\footnote{Herein, the ECS access-cost depends on both the AP and ECS, this is consistent with practical situations. At first, the path from one AP to an ECS may longer than another AP to the same ECS, which then causes a lager delay, and the path from one AP to an ECS is also different with to another ECS \cite{28}. Second, the transmission technology adopted by the backhaul links maybe different among different AP-ECS links \cite{4}, \cite{26}, \cite{27}, \cite{29}, e.g., optical, xDSL or mmWave, et. al. At last, these ECSs and APs may belong to different cloud service providers and virtual network operators (VNOs) \cite{30}, respectively, and then have different agreements on resource using payment \cite{29}.}. In addition, if the task $s_{i,j}$, i.e., the $j$-th ($j=1,...,S_i$) task of MU $a_i~(i=1,...,|\mathcal{A}|)$, is offloaded to any ECS by the AP $b_m$, the delay and energy consumption experienced by task $s_{i,j}$ is characterized by $t_{i,j,m}$ and $e_{i,j,m}$, respectively. The delay and energy consumption costs are different from one user to another, dues to they may have different channel conditions even to the same AP, and also they are different from one task to another, dues to they may have different task parameters, e.g., the amount of offloading task data.

For the APs, we further assume that each of them has a constraint on the allowable offloading connections, i.e., $Q_m$ for AP $b_m$ or the AP $b_m$ at most supports $Q_m$ offloading connections. This constraint can be explained as the maximum available orthogonal communication channels at the AP \cite{12}. For these ECSs, each of them has the computing resource constraint, i.e., $R_n$ for the ECS $c_n$. Following the above, we know that the task offloading involves both the AP association and ECS selection, which can be characterized by the variable $x_{i,j,m,n}$ as
\begin{equation}
x_{i,j,m,n} =
    \begin{cases}
    1& \text{if the task $s_{i,j}$ is offloaded to ECS $c_n$ by AP $b_m$},\\
    0& \text{otherwise}.
    \end{cases}
\label{eq:1}
\end{equation}

Therefore, the offloading cost which including offloading delay, energy consumption and also the ECS access-cost for task $s_{i,j}$ is
\begin{equation}
u_{i,j} = \sum_{m=1}^{|\mathcal{B}|}\sum_{n=1}^{|\mathcal{C}|}x_{i,j,m,n}(\alpha_it_{i,j,m} + \beta_ie_{i,j,m} + \gamma_i\delta_{m,n}).
\label{eq:2}
\end{equation}
Here, $\alpha_i$, $\beta_i$ and $\gamma_i$ denote the weights of the offloading delay, energy consumption and ECS access-cost, respectively, which reflect how MU $a_i$ cares about these overheads in counting the offloading cost. One can note that, user's offloading cost is jointly determined by the accessed AP (delay, energy consumption and ECS access-cost) and ECS (ECS access-cost). This makes the problem difficult to deal with. With the task-level offloading cost definition, the sum offloading cost (of all tasks) for MU $a_i$ can be expressed as
\begin{equation}
U_i = \sum_{j=1}^{S_i}u_{i,j}=\sum_{j=1}^{S_i}\sum_{m=1}^{|\mathcal{B}|}\sum_{n=1}^{|\mathcal{C}|}x_{i,j,m,n}(\alpha_it_{i,j,m} + \beta_ie_{i,j,m} + \gamma_i\delta_{m,n}).
\label{eq:3}
\end{equation}

\subsection{Problem formulations}
With the above, two offloading decision problems with different system objectives, i.e., efficiency and fairness, are formulated. While the efficiency-based objective focuses on minimizing the sum task offloading cost of all MUs. In fact, most of existing works fall into this category \cite{17}-\cite{14}. In specific, for our imbalanced edge cloud system, the problem is characterized as
\begin{equation}
OP1:\min\sum_{i=1}^{|\mathcal{A}|} \sum_{j=1}^{S_i}\sum_{m=1}^{|\mathcal{B}|}\sum_{n=1}^{|\mathcal{C}|}x_{i,j,m,n}(\alpha_it_{i,j,m} + \beta_ie_{i,j,m} + \gamma_i\delta_{m,n})
\label{eq:4}
\end{equation}
\begin{equation}
s.t. \sum_{i=1}^{|\mathcal{A}|}\sum_{j=1}^{S_i}\sum_{m=1}^{|\mathcal{B}|}x_{i,j,m,n}r_{i,j}\leq R_{n},\forall n\in\{1,...,|\mathcal{C}|\},
\label{eq:5}
\end{equation}
\begin{equation}
\sum_{i=1}^{|\mathcal{A}|}\sum_{j=1}^{S_i}\sum_{n=1}^{|\mathcal{C}|}x_{i,j,m,n}\leq Q_{m},\forall m\in\{1,...,|\mathcal{B}|\},
\label{eq:6}
\end{equation}
\begin{equation}
\sum_{m=1}^{|\mathcal{B}|}\sum_{n=1}^{|\mathcal{C}|}x_{i,j,m,n}=1,\forall i\in\{1,...,|\mathcal{A}|\}, j\in\{1,...,S_i\},
\label{eq:7}
\end{equation}
\begin{equation}
x_{i,j,m,n}=\{0,1\},\forall i\in\{1,...,|\mathcal{A}|\}, j\in\{1,...,S_i\}, m\in\{1,...,|\mathcal{B}|\},n\in\{1,...,|\mathcal{C}|\}.
\label{eq:9}
\end{equation}
In which, equation (\ref{eq:5}) guarantees that the total amount of computing resources required by these MUs for the ECS $c_n$ should be less than it has. Equation (\ref{eq:6}) ensures that the connections to an AP do no exceed its available orthogonal channels. Equation (\ref{eq:7}) guarantees that a task of any MU can only and at most be offloaded to one ECS by one AP.

For OP1, with the objective of minimizing the sum task offloading cost of all MUs, even without further analysis but on intuition, we can infer that the resulted solution may cause unfair offloading decision, e.g., some MUs may have small offloading cost for all their tasks, while the others may incur large offloading cost for their tasks. Thus the fairness-based offloading decision problem should be considered, especially for the scenario that each user has multiple offloading tasks. With this in mind, we have the following fairness-based system problem
\begin{equation}
OP2:\min\max_{i}\Bigg\{\eta_i\sum_{j=1}^{S_i}\sum_{m=1}^{|\mathcal{B}|}\sum_{n=1}^{|\mathcal{C}|}x_{i,j,m,n}(\alpha_it_{i,j,m} + \beta_ie_{i,j,m} + \gamma_i\delta_{m,n})/|S_i|\Bigg\},
\label{eq:10}
\end{equation}
\begin{equation}
s.t.~(5),~(6),~(7)~and~(8).
\label{eq:11}
\end{equation}
Herein, we introduce a weight to denote how system cares about the fairness between MUs. In which, $\eta_i$ is the weight allocated to MU $a_i$. Obviously, the larger of $\eta_i$, the smaller offloading cost that MU $a_i$ will have. Comparing OP2 with OP1, the only difference is objective function. In addition, our min-max based fairness is on the user-level but not the task-level, i.e., we ensure that the gaps of the average offloading cost between any two MUs are as small as possible. In fact, to promise task-level fairness is no sense in our system.

\subsection{Feasibility of the problems}
For OP1 and OP2, it is not difficult to infer that their feasible sets maybe empty. Since we have ruled that, all tasks must be offloaded at least by one AP to one ECS, but both the APs and ECSs have self resource constraints. To promise a non-empty feasible set for OP1 or OP2, we implicitly have two additional constraints
\begin{equation}
\sum_{i=1}^{|\mathcal{A}|}\sum_{j=1}^{S_i}r_{i,j}\leq \sum_{n=1}^{|\mathcal{C}|}R_n
\label{eq:12}
\end{equation}
and
\begin{equation}
\sum_{i=1}^{|\mathcal{A}|}\sum_{j=1}^{S_i}\sum_{n=1}^{|\mathcal{C}|}x_{i,j,m,n}\leq \sum_{m=1}^{|\mathcal{B}|}Q_m.
\label{eq:13}
\end{equation}
In which, (\ref{eq:12}) indicates that the total computing resource demands by all MUs is less than that available in the system, and (\ref{eq:13}) limits the number of tasks in the system asking for service. In fact, even (\ref{eq:12}) and (\ref{eq:13}) are satisfied, due to the computing resource requirements from multiple tasks can not align at each ECS, it is still possible that the feasible sets for these problems are empty. This can be explained more clear by the following example even without considering the connection constraints at the APs.

\emph{Example 1}: Suppose that there are two ECS, i.e., $c_1$ and $c_2$, and the available computing resources at these two ECSs are $R_1=3$ and $R_1=4$, respectively. In the system, there are two tasks $s_1$ and $s_2$, whose computing resource requirements are $r_1 =1$ and $r_2 = 5$, respectively. Obviously, we have $r_1+r_2\leq R_1+R_2$ but only one task can be supported, and the problem OP1 has no feasible solution.

To avoid these issues, we assume that the available resources at these APs and ECSs are enough to support all the tasks in the system \cite{12}, \cite{31}, e.g., one of the ECS is a remote cloud server and has enough computing resource, otherwise, an access control or scheduling scheme preceded our algorithm is required. In spite of this, our analysis below shows that both OP1 and OP2 are NP-hard. Therefore, in the sequel, we focus on the suboptimal algorithm design with lower complexity. In fact, it will be more clear that, our following proposed algorithms all have the capability of accessing control.

\section{Efficiency based Problem}
In this section, the efficiency based problem OP1 is discussed and solved. We first analyze its complexity and a linear relaxation algorithm is proposed to characterize the performance bound. Then, both the centralized and distributed suboptimal algorithms are proposed for OP1.

\subsection{The complexity of OP1 and the linear relaxation approach}

At first, we know that OP1 is a combinatorial optimization problem, thus it is NP-hard \cite{12}, \cite{32}, \cite{33}. To obtain its optimal solution, we either need to perform exhaustive search, or derive the sufficient and necessary conditions for its optimal solution. For the former, we have to transverse all possible solutions in the feasible set. However, in our system and for each task, it has $|\mathcal{B}|\times|\mathcal{C}|$ possible offloading pathes, and there are $\sum_{i\in\mathcal{A}}S_i$ tasks in the system, thus the number of offloading path combinations is $(|\mathcal{B}|\times|\mathcal{C}|)^{\sum_{i\in\mathcal{A}}S_i}$, e.g., even for a system has 3 ECSs, 4 APs, 5 MUs with 2 tasks per MU will have $12^{10}$ pathes. That is, even for a small system based problem, its computational complexity is high. For the latter, due to the complicated coupled relationship between AP association and ECS selection, and also the discrete variables, it is impossible for us to obtain the optimal conditions of OP1 .

With the above analysis, a linear relaxation based approach is proposed for OP1, i.e., OP1(LR) as below,
\begin{align*}
OP1(LR):&\min\sum_{i=1}^{|\mathcal{A}|} \sum_{j=1}^{S_i}\sum_{m=1}^{|\mathcal{B}|}\sum_{n=1}^{|\mathcal{C}|}x_{i,j,m,n}\pi_{i,j,m,n}\\
s.t.&~(5),~(6),~(7)~and~x_{i,j,m,n}\in[0,1],\forall i\in\mathcal{A},j\in\mathcal{S}_i,m\in\mathcal{B},n\in\mathcal{C}.\\
\end{align*}
In which, we define $\pi_{i,j,m,n} = \alpha_it_{i,j,m} + \beta_ie_{i,j,m} + \gamma_i\delta_{m,n}$. Note that, now, the binary variables $x_{i,j,m,n}$ are relaxed to real numbers $x_{i,j,m,n}\in[0,1]$. The relaxed problem OP1(LR) is then solved by interior points method (linear relaxation algorithm for efficiency based problem, ELR) which has the complexity in polynomial time $O(v^{3.5}L^2)$, where $v$ is the number of variables and $L$ is the number of the bits in the input \cite{39}. Herein, we have $v=(\sum_{i\in\mathcal{A}}S_i)|\mathcal{B}||\mathcal{C}|$. Denote $\boldsymbol{\hat{X}}$ as the optimal solution of OP1(LR). If all components of $\boldsymbol{\hat{X}}$ are binary, it is also the optimal solution to OP1, otherwise, however, we can not recover binary characteristic of $\boldsymbol{\hat{X}}$ as \cite{13} due to the additional constraints must be satisfied for OP1, i.e., from (\ref{eq:5}) to (\ref{eq:7}). Therefore, the optimal value of OP1(LR) is the lower bound of the problem OP1 and the algorithm ELR can be seen as the performance bound of our following proposed suboptimal algorithms.

\subsection{Centralized greedy algorithm}
Due to the high complexity of directly solve OP1, thus we are interested in designing heuristic algorithms to seek its suboptimal solution. In the sequel, the centralized greedy algorithm (CGA) is proposed. To simplify the presentations, some notations are first introduced below.

$l$: The step index of the algorithm;

$\bar{\mathcal{A}}(l)$: User set with non-empty offloading task set after the $l$th step;

$\bar{\mathcal{S}}^{i}(l)$: Un-offload task set for MU $i\in\bar{\mathcal{A}}(l)$ after the $l$th step;

$Q_m(l)$: The allowable connections for AP $m$ after the $l$th step;

$R_n(l)$: The left computing resource for ECS $n$ after the $l$th step.

$\hat{\mathcal{B}}^{i,j}(l)$: Accessible AP set for task $j\in\bar{\mathcal{S}}^{i}(l)$ of MU $i$ after the $l$th step, and it is defined as
\begin{equation}
\hat{\mathcal{B}}^{i,j}(l) = \{m|Q_m(l)\geq 1,m\in\mathcal{B}\};
\label{eq:14}
\end{equation}

$\hat{\mathcal{C}}^{i,j}(l)$: The accessible ECS set for task $j\in\bar{\mathcal{S}}^{i}(l)$ of MU $i$ after the $l$th step, and it is defined as
\begin{equation}
\hat{\mathcal{C}}^{i,j}(l) = \{n|R_n(l)\geq r_{i,j},n\in\mathcal{C}\};
\label{eq:15}
\end{equation}

$\Delta^{i,j}(l)$: The ECS access-cost matrix for task $j\in\bar{\mathcal{S}}^{i}(l)$ of MU $i$ after the $l$th step, and it is defined as
\begin{equation}
\Delta^{i,j}(l) = \{\delta_{m,n}^{i,j} = \delta_{m,n}|m\in\hat{\mathcal{B}}^{i,j}(l),n\in\hat{\mathcal{C}}^{i,j}(l)\};
\label{eq:16}
\end{equation}

For our proposed algorithm CGA, we assume that the system has a virtual decision center (VDC) who responses for the implementation of the CGA. In particular, at the beginning, VDC collects necessary information for the algorithm, e.g., all tasks' computing resource requirements, their offloading delay and energy consumption over different APs, the ECS access-cost, the connection constraints at the APs and the available computing resources at the ECSs. Then the CGA algorithm is executed at VDC and the results are finally feedbacked to the MUs. The core behind the CGA is the greedy idea as that used in our earlier work \cite{32}, i.e., in each step, the task with the smallest offloading cost is allocated with its optimal offloading path (the AP and ECS which resulted the smallest offloading cost), and the algorithm is end until the left communication or computing resources can not support any more tasks, or all tasks have been offloaded. However, due to the offloading cost of each task is a delay-energy-cost tradeoff, and the (ECS access-) cost depends on both the AP and ECS. Therefore, traditional greedy algorithms used for knapsack problems or weighted bipartite matching \cite{12}, \cite{32}, \cite{33} can not be directly used here. In spite of this, following proposition can be introduced to conduct the algorithm design.

\emph{Proposition 1}: For $\forall i\in\mathcal{A}$ and $\forall j\in\mathcal{S}_i$, if $s_{i,j}$ is offloaded by AP $b_m$, and $\mathcal{C}_m^{i,j} = \{n|R_n\geq r_{i,j}, n\in\mathcal{C}\}$, then the optimal ECS for task $s_{i,j}$ is $n_{o} = \arg\min_{n\in\mathcal{C}_m^{i,j}} \delta_{m,n}$, i.e., the optimal EC is the one has the minimal access-cost with the precondition that its left computing resource is enough for task $s_{i,j}$.

By Proposition 1, a double-layer greedy idea based optimal offloading path selection scheme is proposed for each task and it is the Algorithm 1. The inputs of the algorithm are the accessible AP and ECS sets, and ECS access-cost matrix for task, while the output of the algorithm is the optimal offloading decision, i.e., $(m_o^{i,j},n_o^{i,j},u_{o}^{i,j})$ for task $s_{i,j}$, which characterizes the optimal associated AP, selected ECS and the offloading cost for task $s_{i,j}$.
\begin{algorithm}
\caption{: $(m_o^{i,j},n_o^{i,j},u_{o}^{i,j}) =f(\hat{\mathcal{B}}^{i,j}(l),\hat{\mathcal{C}}^{i,j}(l), \Delta^{i,j}(l))$}
    \begin{algorithmic}[1]
    \STATE Input $\hat{\mathcal{B}}^{i,j}(l)$, $\hat{\mathcal{C}}^{i,j}(l)$ and $\Delta^{i,j}(l)$;

    \STATE According to $\Delta^{i,j}(l)$, search $\hat{n}_{m}^{i,j} = \arg\min_{n\in\hat{\mathcal{C}}^{i,j}(l)}\delta_{m,n}^{i,j}$ for $\forall j\in\bar{\mathcal{S}}^i(l)$;

    \STATE Calculate $m_{o}^{i,j} = \arg\min_{m\in\hat{\mathcal{B}}^{i,j}(l)}{u_m^{i,j}=\alpha_i t_{i,j,m} + \beta_i e_{i,j,m} + \gamma_i \delta_{m,\hat{n}_{m}^{i,j}}^{i,j}}$, and let $n_{o}^{i,j} = \hat{n}_{m_{o}^{i,j}}^{i,j}$ and $u_{o}^{i,j} = u_{m_{o}^{i,j}}^{i,j}$;

    \STATE Output $(m_o^{i,j},n_o^{i,j},u_{o}^{i,j})$.
    \end{algorithmic}
\end{algorithm}

With the optimal offloading path decision algorithm for each task, we then present the algorithm CGA for OP1 which is summarized in the Algorithm 2. At first, at the $l$th step, according to the available communication and computing resource, the accessible AP and ECS sets, the ECS access-cost matrix for each un-offload task is updated, i.e., $\hat{\mathcal{B}}^{i,j}(l)$, $\hat{\mathcal{C}}^{i,j}(l)$ and $\Delta^{i,j}(l)$ for task $s_{i,j}$; then, for each un-offload task, the optimal offloading path is searched, i.e., we then have $(m_o^{i,j},n_o^{i,j},u_{o}^{i,j})$ for task $s_{i,j}$; in the sequel, for each MU $ i\in\bar{\mathcal{A}}(l)$, calculates the task $j_o$ with the minimum offloading cost, i.e., the step 2-3); after that, the comparison is done between different MUs in $\bar{\mathcal{A}}(l)$ to obtain the task with the smallest offloading task, i.e., the task $j^\star$ of $i^\star$, and it is offloaded with its optimal offloading path, i.e., it is offloaded by AP $m^\star$ to ECS $n^\star$; finally, both the left computing and communication resources, and the un-offload task set for MU $i^\star$ are updated. The algorithm is end when one of the following three conditions becomes true: 1) the left computing resource can not support any more task, i.e., $\max_{n\in\mathcal{C}}R_n(l)\leq\min_{i\in\bar{\mathcal{A}}(l),j\in\bar{\mathcal{S}}^i(l)}r_{i,j}$; 2) the left communication resource is not enough to support any connections, i.e., $Q_m(l)<1,m\in\mathcal{B}$; 3) all tasks have been offloaded, i.e., $\bar{\mathcal{A}}(l)=\Phi$.

\begin{algorithm}
\caption{: Centralized greedy algorithm (CGA)}
    \begin{algorithmic}[1]
        \STATE Initialize $l=0$, $Q_m(l)=Q_m$, $R_n(l) = R_n$, $\bar{\mathcal{A}}(l)=\mathcal{A}$ and $\bar{\mathcal{S}}^{i}(l) = \mathcal{S}_{i}$;
        \STATE For $\forall i\in\bar{\mathcal{A}}(l)$, successively implements the following steps:

        \begin{enumerate}
            \item Update $\hat{\mathcal{B}}^{i,j}(l)$, $\hat{\mathcal{C}}^{i,j}(l)$ and $\Delta^{i,j}(l)$ for $\forall j\in\bar{\mathcal{S}}^i(l)$ by (\ref{eq:14}), (\ref{eq:15}) and (\ref{eq:16}), respectively;

            \item Perform \textbf{Algorithm 1} for $\forall j\in\bar{\mathcal{S}}^i(l)$, i.e., $(m_o^{i,j},n_o^{i,j},u_{o}^{i,j})=f(\hat{\mathcal{B}}^{i,j}(l),\hat{\mathcal{C}}^{i,j}(l), \Delta^{i,j}(l))$;

            \item Calculate $j_{o}^i = \arg\min_{j\in\bar{\mathcal{S}}^{i}(l)}u_{o}^{i,j}$, and let $u_{o}^i = u_{o}^{i,j_{o}^{i}}$, $m_{o}^i = m_{o}^{i,j_{o}^{i}}$ and $n_{o}^i = n_{o}^{i,j_{o}^{i}}$;
        \end{enumerate}

        \STATE Calculate $i^\star = \arg\min_{i\in\bar{\mathcal{A}}(l)}u_{o}^i$, and let $j^\star = j_{o}^{i^\star}$, $m^\star = m_{o}^{i^\star}$, $n^\star = n_{o}^{i^\star}$ and $u^\star =u_{o}^{i^\star}$, i.e., the task $j^\star$ of MU $i^\star$ is offloaded from AP $m^\star$ to ECS $n^\star$;

        \STATE Update $R_{n^\star}(l) = R_{n^\star}(l) - r_{i^\star,j^\star}$, $Q_{m^\star}(l) = Q_{m^\star}(l) - 1$ and $\bar{\mathcal{S}}^{i^\star}(l) = \bar{\mathcal{S}}^{i^\star}(l)\setminus j^\star$; if $\bar{\mathcal{S}}^{i^\star}(l) =\Phi$, $\bar{\mathcal{A}}(l)=\bar{\mathcal{A}}(l)\setminus i^\star$;

        \STATE If $\max_{n\in\mathcal{C}}R_n(l)\leq\min_{i\in\bar{\mathcal{A}}(l),j\in\bar{\mathcal{S}}^i(l)}r_{i,j}$, or $Q_m(l)<1,m\in\mathcal{B}$, or $\bar{\mathcal{A}}(l)=\Phi$, the algorithm is end; otherwise, $l = l +1$, goto step 2.
    \end{algorithmic}
\end{algorithm}

Since the convergence of the CGA is obvious, thus we only discuss its complexity and have the conclusion below.

\emph{Proposition 2}: The algorithm complexity of the CGA is upper bound by $O(\sum_{i\in\mathcal{A}}|\mathcal{S}_i|(|\mathcal{A}|^2 + |\mathcal{A}|(|\mathcal{S}_i|^2 +|\mathcal{S}_i|(|\mathcal{B}|^2+|\mathcal{C}|^2)))+(\sum_{i\in\mathcal{A}}|\mathcal{S}_i|)^2+|\mathcal{C}|^2)$\footnote{In this paper, since the computing operations involved in the algorithm are simple, thus the algorithm complexity only counts the number of sort operations in the algorithm who is more computing intensive, i.e., for a vector $\bf{v}$ with $N$ elements, the sort complexity of $\bf{v}$ is $O(N^2)$.}.

\textit{Proof:} Firstly, for the algorithm CGA, we know that at each step, we make an offloading decision for one task, and the algorithm is end whenever the left communication or computing resources can not support any more tasks, or all tasks have been offloaded. Thus, the algorithm will be end after no more than $\sum_{i\in\mathcal{A}}|\mathcal{S}_i|$ iterations.

Second, at each iteration of the algorithm CGA, the computing intensive operations are the sort operations in the step 2-2), step 2-3), step 3 and step 5. For the step 2-2), i.e., the Algorithm 1, it is performed at task-level and for each task $j\in\bar{\mathcal{S}^i}(l)$, the sort operations have the complexity of $O(|\mathcal{C}|^2+|\mathcal{B}|^2)$; for the step 2-3), it is performed at the user-level and its complexity is at most $O(|\mathcal{S}_i|^2)$; for the step 3, it is performed at the system-level and has the complexity of $O(|\mathcal{A}|^2)$; for the step 5, though it seems that we need check the condition $\max_{n\in\mathcal{C}}R_n(l)\leq\min_{i\in\bar{\mathcal{A}}(l),j\in\bar{\mathcal{S}}^i(l)}r_{i,j}$ at each iteration, in fact, we only need to calculate it one time and then update its parameters each step with low complexity, and the complexity is $(\sum_{i\in\mathcal{A}}|\mathcal{S}_i|)^2+|\mathcal{C}|^2$.

To sum up, the complexity of the algorithm CGA is $O(\sum_{i\in\mathcal{A}}|\mathcal{S}_i|(|\mathcal{A}|^2 + |\mathcal{A}|(|\mathcal{S}_i|^2 +|\mathcal{S}_i|(|\mathcal{B}|^2+|\mathcal{C}|^2)))+(\sum_{i\in\mathcal{A}}|\mathcal{S}_i|)^2+|\mathcal{C}|^2)$. $\Box$

Before concluding this subsection, we present an example to explain why the algorithm CGA is suboptimal. Then we propose a modified greedy algorithm (MGA), it is still a centralized algorithm and thus an VDC is required too. Though theoretically proving its optimality over CGA is impossible, we can infer it will superior to the CGA at some circumstances, e.g., as the example below. In addition, simulation results will be presented in Section V to verify this statement.

\emph{Example 2}: Suppose there are two ECSs, e.g., $c_1$ and $c_2$, and their computing resources are $R_1$ and $R_2$, respectively. In the system, there are three offloadable tasks, e.g., $s_1$, $s_2$ and $s_3$, their computing resource requirements are $r_1$, $r_2$ and $r_3$, respectively. In addition, we assume that
\begin{align*}
r_1\leq R_1,~r_1\leq R_2,~R_1-r_1<r_2,~R_1-r_1<r_3,\\
r_2+r_3\leq R_1,~r_2+r_3\leq R_2,~R_1-(r_2+r_3)<r_1.
\end{align*}
That is, both $c_1$ and $c_2$ can support $s_1$, $s_2$ and $s_3$. However, if $c_1$ is allocated to $s_1$, it can neither support $s_2$ nor $s_3$, if $c_1$ is allocated to $s_2$ and $s_3$, it then can not support $s_1$ any more. For these three tasks, i.e., $s_1$, $s_2$ and $s_3$, assuming their offloading costs to $c_1$ and $c_2$ are $u_1^1$, $u_1^2$, $u_2^1$, $u_2^2$, $u_3^1$ and $u_3^2$, respectively. We further assume that
\begin{align*}
u_1^1<u_1^2,~u_2^1<u_2^2,~u_3^1<u_3^2,~u_1^1<u_2^1,~u_1^1<u_3^1.
\end{align*}
Then for the CGA, its offloading decisions are, $s_1$ is allocated to $c_1$, and $s_2$ and $s_3$ are allocated with $c_2$, obviously, it is a feasible solution and its total offloading cost is $u_{CGA} = u_1^1+u_2^2+u_3^2$. However, another feasible solution is that, $s_1$ is allocated with $c_2$, and $s_2$ and $s_3$ are allocated with $c_1$, and the resulted total offloading cost is $u' = u_2^1+u_3^1 + u_1^2$. In particular, if
\begin{align*}
(u_2^2+u_3^2) - (u_2^1+u_3^1)>(u_1^2-u_1^1)>0,
\end{align*}
we have $u'<u_{CGA}$ which verified the suboptimality of the CGA.

By observing the Example 2, one can note that for the CGA, we only consider the offloading cost in making an offloading decision but ignore the effect of computing resource consumption. This of course results a sub-optimality solution. To handle this issue, we introduce a new metric in making offloading decision, i.e., for task $s_{i,j}$, we define a modified offloading cost as
\begin{equation}
\hat{u}_{i,j} = u_{i,j}^{\epsilon}r_{i,j}^{\zeta}.
\label{eq:17}
\end{equation}
That is, not only the offloading cost but also the computing resource requirement are included in the modified offloading cost. In which, $\epsilon\geq 1$ and $\zeta\geq 1$ are two parameters used to weight the effects of original offloading cost and the computing resource consumption. Based on (\ref{eq:17}), the step 2-3) of the CGA can be modified as $j_{o}^i = \arg\min_{j\in\bar{\mathcal{S}}^{i}(l)}\hat{u}_{o}^{i,j}$, in which we use $\hat{u}_{i,j}$ to replace $u_{i,j}$. The resulted algorithm is the modified greedy algorithm (MGA). For (\ref{eq:17}), in fact, the ratio of $\epsilon$ and $\zeta$ but not their values is important, and which is used to balance the effects of offloading cost and resource consumption in making offloading decision. However, theoretically deriving the optimal value of these two parameters is impossible, thus we only use simulations to analyze their effects on the performance of the algorithm.

\subsection{Matching based distributed task offloading}
The motivation of the work in this subsection comes from two shortcomings of the CGA. On the one hand, at each step of the CGA, only one task offloading decision is made, thus it is time-consuming and with poor scalability, especially when there are lots of tasks in the system but the communication and computing resources are sufficient. On the other hand, the CGA is a centralized algorithm which requires an VDC to collect the necessary information from all MUs, APs and ECSs. However, a centralized VDC may not exist in the system. In order to accelerate the task offloading decision and relax the requirement of an VDC, by relaxing OP1, i.e., we remove the connection constraint ($Q_m, m\in\mathcal{B}$) at the AP and assume all tasks in the system have the same amount of computing resource requirements \cite{12}, i.e., $r_{i,j} =r,\forall i\in\mathcal{A},j\in\mathcal{B}$ and based on the many-to-one matching game \cite{34}-\cite{37}, an asynchronous distributed matching algorithm (ADMA) is presented herein.

\subsubsection{A primer on matching game}
Matching game provides mathematically tractable solutions for the combinatorial problem of matching players in two distinct sets, which depending on the preference of each player. Leveraging the one-to-one stable marriage problem \cite{34}, \cite{35}, we first introduce the theory of matching game.

\emph{Stable marriage problem}: Considering two disjoint sets, $\mathcal{M}=\{m_1,m_2,...,m_Q\}$ and $\mathcal{W} = \{w_1,w_2,...,w_P\}$ which denote the set of different agents, e.g., men and women, respectively. The agent in $\mathcal{M}$ has a transitive preference over individuals on the other side, i.e., $\mathcal{W}$, and also may prefer to be unmatched. In general, we use $\succ_i$ to denote the ordering relationship of $i$, e.g., for $m_1$, the ordering relationship is $w_3\succ_{m_1}w_4\succ_{m_1},...$, which means that the first choice of $m_1$ is $w_3$, and the second choice is $w_4$ and so on, and $\phi\succ_{m_1}w_j$ indicates that $w_j$ is unacceptable to $m_1$. In mathematics, an one-to-one matching game can be rigorously defined as below.

\emph{Definition 1}: A matching is a mapping $\mu:\mathcal{M}\times\mathcal{W}\times\Phi\to \mathcal{M}\times\mathcal{W}\times\Phi$ such that $w=\mu(m)$ if and only if $\mu(w) = m$, and $\mu(m)\in\mathcal{W}\bigcup\Phi$, $\mu(w)\in\mathcal{M}\bigcup\Phi$, $\forall m,w$.

In order to characterize the outcome of the matching, stable matching is introduced and it is the mostly used solution concept in the matching problem.

\emph{Definition 2}: A matching $\mu$ is individual rational to all agents (i.e., men and women), if and only if there does not exist an agent $i$ who prefers being unmatched to being matched with $\mu(i)$, i.e., $\phi\succ_{i}\mu(i)$.

\emph{Definition 3}: A matching $\mu$ is blocked by a pair of agents $(m,w)$ if they each prefer each other to the partner they receive at $\mu$. That is, $w\succ_m\mu(m)$ and $m\succ_w\mu(w)$. Such a pair is called a blocking pair in general.

\emph{Definition 4}: A matching $\mu$ is stable if and only if it is individual rational, and not blocked by any pair of agents.

For the stable marriage problem, Gale and Shapley have already proved that a stable matching is always exist \cite{33}. Though it's simple and it is an one-to-one matching, there are multiple extensions, e.g., many-to-one matching, matching with incomplete preference lists or tier \cite{36}, \cite{37}. For the many-to-one matching, each agent in one side can match with multiple agents on the other side, e.g., the task offloading matching discussed below.

\subsubsection{Task offloading matching}
The task offloading decision problem which involves both AP association and ECS selection is modeled as the many-to-one task offloading matching (TOM) problem as below.

\begin{description}
  \item[a.] In the TOM, the task is treated as an independent player but not the MU, i.e., the tasks in $\mathcal{S} =\{s_{i,j}|i\in\mathcal{A},j\in\mathcal{S}_i\}$ formulate one side players of the matching game. Each player aims at AP association and ECS selection to achieve task offloading with the objective of minimizing the offloading cost. We know that a task will incur diverse offloading cost if it is offloaded by different AP or ECS, thus task has preference over all possible offloading paths. In addition, in our system, each task can select only one offloading path, i.e., associating with one AP and selecting one ECS.

  \item[b.] The ECSs but not the APs formulate the other side players of the matching game. In general, the available computing resource at each ECS can support multi-tasks, i.e., one ECS matches multiple tasks, thus the TOM is a many-to-one matching. In particular, for ECS $n$, it can support at most $w_n =\lfloor R_n/r \rfloor$ tasks.

  \item[c.] In addition, we rule that the tasks are the proposers, i.e., the tasks propose the matching request and then the ECS makes the decision whether or not to accept the request.
\end{description}

For the above TOM, there are three important issues should be deal with. At first, originally, the ECS has no preference over different tasks. However, on the one hand, each ECS can match multiple tasks. On the other hand, ECS has a constraint on the available computing resource, thus the resource request may overflow and this is problematic. To tackle this issue and also consider the system objective defined in OP1, the task offloading cost based preference rule (OCPR) for ECS is introduced\footnote{The preference rule of ECS is not unique and depended on the system design objective, thus some other criterions may still valid, e.g., the amount of request computing resource.}, then we propose the OCPR based matching scheme for ECS.

\emph{Task offloading cost based preference rule (OCPR) for ECS}: For the OCPR, ECS's preference over different tasks is ranked by the task's offloading cost, i.e., for ECS $n\in\mathcal{C}$ and its service asking task set $\mathcal{V}_n=\{s_{n}^{z}| s_{n}^{z}=s_{i,j},~if~s_{i,j}\in\mathcal{S}~and~n_o^{i,j}=n, z=1,...,|\mathcal{V}_n|\}$, we have $\forall s_{n}^{z},s_{n}^{z'}\in\mathcal{V}_n$, $s_{n}^{z}\succ_{c_n}s_{n}^{z'}$ if and only if $u(s_{n}^{z})\leq u(s_{n}^{z'})$, in which $n_o^{i,j}=n$ indicates that the task $s_{i,j}$ chooses the $n$th ECS as the service ECS in the offloading decision.

\emph{OCPR based matching scheme for the ECS}: As mentioned earlier, the computing resource requests from multiple tasks may lead to an overflow at the ECS, thus the ECS should perform the matching task selection under some schemes. For the OCRP based matching scheme, ECS first ranks the tasks in $\mathcal{V}_n$ by OCPR on an ascending order. Then based on the computing resource constraint or the number of supportable tasks, i.e., $w_n$, ECS performs the matching task selection. The whole process is summarized in the Algorithm 3.
\begin{algorithm}
\caption{: $(\hat{\mathcal{V}}_n, \bar{\mathcal{V}}_n)=f(w_n,\mathcal{V}_n)$}
    \begin{algorithmic}[1]
    \STATE Input $w_n$ and $\mathcal{V}_n=\{s_n^z,z=1,...,|\mathcal{V}_n|\}$;

    \STATE Initialize $\tilde{\mathcal{V}}_n=\Phi$, $\hat{\mathcal{V}}_n=\Phi$ and $\bar{\mathcal{V}}_n=\Phi$;

    \STATE If $|\mathcal{V}_n|>w_n$, goto step 4; otherwise, let $\hat{\mathcal{V}}_n =\mathcal{V}_n$ and goto step 5;

    \STATE According to the OCPR, ECS ranks the tasks in $\mathcal{V}_n$, then we have $\tilde{\mathcal{V}}_n =\{s_n^{[z]},z=1,...,|\mathcal{V}_n|\}$, let $\hat{\mathcal{V}}_n=\{s_n^{[z]},z=1,...,w_n\}$ and $\bar{\mathcal{V}}_n= \{s_n^{[z]},z=w_n+1,...,|\mathcal{V}_n|\}$;

    \STATE Output $\hat{\mathcal{V}}_n$ and $\bar{\mathcal{V}}_n$.
    \end{algorithmic}
\end{algorithm}

\begin{figure}[]
\centering
\includegraphics[width=3.5in]{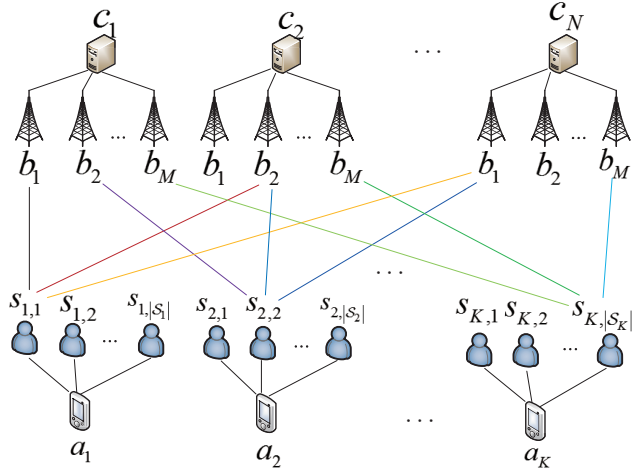}
\caption{Task offloading matching (TOM) model.}
\label{fig.3}
\end{figure}

Second, though the APs do not serve as the other side players in the matching game, they have impacts on the the tasks' offloading decision making. This issue can be explained by the TOM model as shown in the Fig. 2. In which, each ECS is connected with multiple APs\footnote{As mentioned earlier, we assume that each ECS is connected and shared by all the APs, and this assumption can be relaxed for the scenario that only a subset of APs connected to the ECS.}. For a task, even for given ECS selection, it still needs to determine which AP to associate with. With different AP association, the task will incur different offloading delay and energy consumption. In order to simplify the illustrations, let $p_{m,n}^{i,j}$ denote the offloading path from AP $m$ to ECS $n$ for task $s_{i,j}$, then the strategy set $\mathcal{P}^{i,j}$ for the task $s_{i,j}$ is the all possible offloading paths, i.e.,
\begin{equation}
\mathcal{P}^{i,j} = \{p_{m,n}^{i,j}|m\in\hat{\mathcal{B}}^{i,j},n\in\hat{\mathcal{C}}^{i,j}\}.
\label{eq:21}
\end{equation}
In which, since we have relaxed the connection constraints at the APs, then $\hat{\mathcal{B}}^{i,j}=\mathcal{B}$ and $\hat{\mathcal{C}}^{i,j}$ is defined in (\ref{eq:15}) by letting $l=0$.
From (\ref{eq:21}), it is not difficult to infer that $|\mathcal{P}^{i,j}| = |\mathcal{B}|\times|\hat{\mathcal{C}}^{i,j}|$. However, the proposition below states that the volume of $\mathcal{P}^{i,j}$ can be significantly reduced.

\emph{Proposition 3}: If the OCPR based matching scheme is used by the ECS, i.e., the Algorithm 3, the strategy set for task $s_{i,j}$ is
\begin{equation}
\hat{\mathcal{P}}^{i,j} =\{p_{\hat{m},n}^{i,j}|n\in\hat{\mathcal{C}}^{i,j},\hat{m} = \arg\min_{m\in\mathcal{B}}u(p_{m,n}^{i,j})\},
\label{eq:19}
\end{equation}
and $|\hat{\mathcal{P}}^{i,j}| = |\hat{\mathcal{C}}^{i,j}|$, i.e., the number of available strategies for task $s_{i,j}$ will not exceed $|\mathcal{C}|$.

\textit{Proof:} At first, defining $\mathcal{P}_n^{i,j}$ as the accessible AP set for task $s_{i,j}$ if ECS $n$ is selected as its offloading server, then $\mathcal{P}^{i,j} =\bigcup_{n\in\mathcal{C}}\mathcal{P}_n^{i,j}$. In fact, $\mathcal{P}_n^{i,j}=\mathcal{B}$.
Obviously, we have
\begin{equation}
u(p_{\hat{m},n}^{i,j})\leq u(p_{m,n}^{i,j}),\forall m\in\mathcal{P}_n^{i,j}.
\label{eq:20}
\end{equation}

Second, we define $\mathcal{V}_n=\{s_n^z,z=1,...,|\mathcal{V}_n|-1\}\bigcup\{s_{i,j}\}$ as the service asking task set for ECS $n$. Then based on the OCPR matching scheme, ECS ranks the tasks in $\mathcal{V}_n$ to perform task selection, we have
\begin{equation}
s_n^{[1]}\succ_{c_n}s_n^{[2]}\succ_{c_n}...s_n^{[\hat{z}]}\succ_{c_n}s_{i,j}\succ_{c_n}s_n^{[\hat{z}+1]}\succ_{c_n}...\succ_{c_n}s_n^{[|\mathcal{V}_n|]}
\label{eq:21}
\end{equation}
and
\begin{equation}
u(s_n^{[1]})\leq u(s_n^{[2]})\leq...u(s_n^{[\hat{z}]})\leq u(p_{\hat{m},n}^{i,j})\leq u(s_n^{[\hat{z}+1]})\leq...\leq u(s_n^{[|\mathcal{V}_n|]}).
\label{eq:22}
\end{equation}
Since we have assumed that $r=r_{i,j},\forall i\in\mathcal{A},j\in\mathcal{B}$, then two cases may happen for $s_{i,j}$: 1) $(\hat{z}+1)r\leq R_n$, then at least $\hat{z}+1$ tasks can be supported by the ECS $n$ and task $s_{i,j}$ will be matched by the ECS; 2) $(\hat{z}+1)r > R_n$, then the ECS $n$ only can support less than $\hat{z}$ tasks and task $s_{i,j}$'s service request will be rejected. In addition, once the task $s_{i,j}$ is refused by the ECS $n$, due to the relationship of (\ref{eq:20}), its service request from any other AP $m'\in\mathcal{P}_n^{i,j}$ to ECS $n$ will be rejected also. That is, for each ECS $n$, task $s_{i,j}$ has one choice on AP association, i.e., the $\hat{m}\in\mathcal{P}_n^{i,j}$ resulted the smallest offloading cost. $\Box$

By the Proposition 3, the jointly AP association and ECS selection is degrade to only the ECS selection, and the associated AP is the one who results the least offloading cost under current ECS selection.

At last, since we assume that all MUs can not support the locally task executing and all of them should be offloaded to an ECS, also the system has enough communication and computing resource. Therefore, for the TOM problem, a task at least and at most match one ECS, however, the ECS may not match any task. This makes our TOM problem not the same as traditional matching problem and then causes different matching solution concept, i.e., all the players in $\mathcal{S}$ are matched but not the same case for ECS, also we do not need to consider the individual rational. In particular, we formally define the task offloading matching as follows.

\emph{Definition 5 (Task offloading matching: TOM)}: A matching $\mu$ is a mapping from $\mathcal{S}\bigcup\mathcal{C}$ to $\mathcal{S}\bigcup\mathcal{C}$ which satisfies:

i. $\mu(s_{i,j})\in\mathcal{C}$ and $|\mu(s_{i,j})|\leq 1$ if $s_{i,j}\in\mathcal{S}$,

ii. $\mu(c_n)\subseteq\mathcal{S}\bigcup c_n$ and $\sum_{\mu(s_{i,j})=c_n}r_{i,j}\leq R_n$ if $c_n\in\mathcal{C}$,

iii. $s_{i,j}\in\mu(c_n)$ if and only if $\mu(s_{i,j})=c_n$.
\newline In which, (i) indicate that task $s_{i,j}$ at most matches with one AP and the matched AP is $\mu(s_{i,j})$; (ii) and indicate that an AP can match multiple tasks or itself and for the former, it has the additional constraint that the asking computing resources can not exceed it has; (iii) ensures symmetry in the matching.

Now, we introduce the solution concepts for the TOM, i.e., a matching blocked by a pair and a stable matching.

\emph{Definition 6}: For TOM $\mu$, $(s_{i,j},c_n)$ is a blocking pair if one of the following conditions hold:

i.  $c_n\succ_{s_{i,j}}\mu(s_{i,j})$, if $(|\mu(c_n)|+1)r\leq R_n$,

ii. $c_n\succ_{s_{i,j}}\mu(s_{i,j})$ and $s_{i,j}\succ_{c_n}s_n^{[|\mu(c_n)|]}$if $(|\mu(c_n)|+1)r> R_n$.
\newline In which, $s_n^{[|\mu(c_n)|]}$ is the task with the largest offloading cost in $\mathcal{V}_n$. The first condition indicates that under matching $\mu$, the ECS $n$ at leat can support one more task, and task $s_{i,j}$ will have less offloading cost when match ECS $n$ than its currently matched ECS $\mu(s_{i,j})$; the second condition means that under matching $\mu$, the ECS $n$ can not support any more tasks, however, task $s_{i,j}$ will have less offloading cost when match ECS $n$ than its currently matched ECS $\mu(s_{i,j})$, in addition, its offloading cost is less than the one with the largest offloading cost in $\mu(c_n)$, i.e., $s_n^{[|\mu(c_n)|]}$.

\emph{Definition 7}: A TOM is stable if it does not contain any blocking pair.
\begin{algorithm}
\caption{: Asynchronous distributed matching algorithm (ADMA)}
    \begin{algorithmic}[1]
        \STATE Initialize $l=0$, $\bar{\mathcal{A}}(l)=\mathcal{A}$, $\bar{\mathcal{S}}^{i}(l) = \mathcal{S}_{i}$ and $\hat{\mathcal{S}}^{i}(l) = \Phi$;

        \STATE For $\forall j\in\bar{\mathcal{S}}^{i}(l)$ and $\forall i\in\bar{\mathcal{A}}(l)$, initialize $\hat{\mathcal{B}}^{i,j}(l) = \mathcal{B}$ and $\hat{\mathcal{C}}^{i,j}(l)$ by (\ref{eq:15});

        \STATE For $\forall j\in\bar{\mathcal{S}}^{i}(l)$ and $\forall i\in\bar{\mathcal{A}}(l)$, implements the following steps:
        \begin{enumerate}
            \item According to $\hat{\mathcal{B}}^{i,j}(l)$ and $\hat{\mathcal{C}}^{i,j}(l)$, updating $\Delta^{i,j}(l)$ by (\ref{eq:16});

            \item Performs \textbf{Algorithm 1} for $\forall j\in\bar{\mathcal{S}}^i(l)$, i.e., $(m_o^{i,j},n_o^{i,j},u_{o}^{i,j})=f(\hat{\mathcal{B}}^{i,j}(l),\hat{\mathcal{C}}^{i,j}(l), \Delta^{i,j}(l))$;

            \item MU $i$ sends the offloading decision message of task $j$ which includes $(r_{i,j},u_{o}^{i,j})$ to ECS $n_{o}^{i,j}$;
        \end{enumerate}

        \STATE For ECS $n$ and its service asking task set $\mathcal{V}_n(l)=\{s_n^z, z=1,...,|\mathcal{V}_n|\}$, performs the following steps:
        \begin{enumerate}
            \item Implements the \textbf{Algorithm 3}, i.e., $(\hat{\mathcal{V}}_n, \bar{\mathcal{V}}_n)=f(w_n,\mathcal{V}_n)$;

            \item If $\bar{\mathcal{V}}_n\neq\Phi$, ECS $n$ sends the service rejection message to tasks belonged to the task set $\bar{\mathcal{V}}_n$;
        \end{enumerate}

        \STATE For $\forall j\in\bar{\mathcal{S}}^{i}(l)$ and $\forall i\in\bar{\mathcal{A}}(l)$, if its service request is rejected by ECS $n$, update $\bar{\mathcal{C}}^{i,j}(l)=\bar{\mathcal{C}}^{i,j}(l)\setminus n$, $\bar{\mathcal{S}}^{i}(l) = \bar{\mathcal{S}}^{i}(l)\bigcup j$ and $\hat{\mathcal{S}}^{i}(l) = \hat{\mathcal{S}}^{i}(l)\setminus j$, otherwise, update $\bar{\mathcal{S}}^{i}(l) = \bar{\mathcal{S}}^{i}(l)\setminus j$ and $\hat{\mathcal{S}}^{i}(l) = \hat{\mathcal{S}}^{i}(l)\bigcup j$;

        \STATE If $\bigcup_{i\in\mathcal{A}}\bar{\mathcal{S}}^i(l) = \Phi$, or $\bigcup_{i\in\mathcal{A},j\in\bar{\mathcal{S}}^i(l)}\hat{\mathcal{C}}^{i,j}(l) =\Phi$, the algorithm is end; otherwise, $l = l +1$, goto step 3.
    \end{algorithmic}
\end{algorithm}
\subsubsection{Asynchronous distributed matching algorithm (ADMA)}
Following the prementioned TOM model, an asynchronous distributed matching algorithm is proposed for the relaxed OP1. In which, $\mathcal{V}_n(l)=\{s_{i,j}|i\in\bar{\mathcal{A}},j\in\bar{\mathcal{S}}_i\}$ denotes the service asking task set for ECS $n$ in the $l$th step. At first, we initialize the system status, i.e., the offloaded and un-offload task set, et. al. Then, for MU and ECS, we separately and asynchronously perform the left steps until the system is stable.

For each MU, we independently constructs the accessible AP and ECS sets for all its un-offload tasks, i.e., $\hat{\mathcal{B}}^{i,j}(l)$ and $\hat{\mathcal{C}}^{i,j}(l)$, then the optimal offloading path for each un-offload task is calculated. The result is formulated as a service request and sent to the corresponding ECS; once this request is rejected by the ECS, then the task updates its accessible ECS set and repeats the above processes until no more rejections for its request, or its accessible ECS set is empty. For the ECS, whenever received a service request, it will check whether the asking computing resources exceed it has. If so, according to its matching rule, the ECS rejects the task(s) with the largest offloading cost(s) who causes the overflow. The details of the algorithm are summarized in Algorithm 4.
\subsubsection{Algorithm analysis}
For the algorithm ADMA, since each MU (or task) independently makes an offloading decision, thus it is a distributed algorithm. In addition, there is no need of synchronization between MU and ECS for its implementation, i.e., an ECS can immediately send the service refusing message whenever the resource request exceed it has, and an MU can independently send a new service request whenever its pre-service request has been rejected. Some other properties of the algorithm ADMA are summarized in the following proposition.

\emph{Proposition 4}: The algorithm ADMA will converge after no more than $(\sum_{i\in\mathcal{A}}S_i)|\mathcal{C}|$ iterations and finally converge to a stable matching.

\textit{Proof:} This conclusion can be directly derived from the Proposition 3 and the process of the Algorithm 4.
$\Box$

To conclude with this section, it is worth pointing out that, the ADMA algorithm still work even the asked computing resources from different tasks are not equal. However, in this case, two issues exist for this algorithm. At first, since we can not align the task's offloading cost and resource requirement, then the performance of sum offloading cost may increase, i.e., there are some tasks who have small offloading cost but with large computing resource requirements. Second, the outcome is not a stable matching, i.e., some tasks may obtain lower offloading cost if it is matched with an ECS who has already rejected him/her than his/her currently matched. These problems can be explained by the following example.

\emph{Example 3}: We assume there are two ECS in the system, i.e., $c_1$ and $c_2$, there are six tasks in the system, i.e., $s_i,i=1,...,6$. As that in the Example 2, we ignore the effects from APs. Assuming $c_1$ and $c_2$'s computing resources are $R_1=10$ and $R_2=12$, respectively. Suppose that at the beginning, the tasks accessed into ECS $c_1$ and $c_2$ are $\mathcal{S}_1=\{s_1, s_3\}$ and $\mathcal{S}_1=\{s_2, s_4, s_5\}$, respectively. Their offloading costs and computing resource requirements are shown in the first row and second row of Table 1, respectively. From the model, we know that $r_1+r_3>R_1$ and $r_2+r_4+r_5<R_2$, i.e., $c_1$ can not simultaneously support $s_1$ and $s_3$, then following the OCPR, $r_3$ will be rejected by ECS $c_1$, however, $c_2$ can simultaneously support $s_2$, $s_4$ and $s_5$. We assume that the rejected task $s_3$ can access into $c_2$ with the same offloading cost. Thus, it will send service request to ECS $c_2$. However, since $r_2+r_3+r_4+r_5>R_2$, then following the OCPR, $s_4$ and $s_5$ will be rejected by ECS $c_2$. In this case, if there are another task $s_6$ wants to access into ECS $c_2$ with the offloading cost 2.5 and computing resource $3$, we know that $r_3$ will be rejected by ECS $c_2$. Then the final result of this matching is, ECS $c_1$ matches $s_1$, and ECS $c_s$ matches $s_2$ and $s_6$. However, we know the rejected tasks $s_4$ and $s_5$ in fact can access into ECS $c_2$ and without violate its resource constraint, thus the matching is not stable.
\begin{table}
\centering
\caption{$~$Offloading cost and computing resource requirements for six tasks}
\begin{tabular}{ccccccc}
\hline
   & $s_1$ & $s_2$ & $s_3$ & $s_4$ & $s_5$ & $s_6$\\
\hline
$u_i$ &  1  &  2  &  3  &  4  &  5  &  2.5\\
$r_i$ &  1  &  2  &  10 &  3  &  1  &  3\\
\hline
\end{tabular}
\end{table}

\section{Fairness based Problem}
In this section, the fairness based problem OP2 is discussed and solved. As mentioned earlier, since the OP2 is a combinational optimization problem and intractable. Therefore, our focus is on the suboptimal algorithm design. However, at first, a linear relaxation based performance bound algorithm is presented for OP2. In specific, by introducing a variable $y$, then OP2 can be transformed into an equivalent problem OP3 as follows,
\begin{equation}
OP3:\min_{x_{i,j,m,n},y} y
\label{eq:29}
\end{equation}
\begin{equation}
s.t.\sum_{j=1}^{S_i}\sum_{m=1}^{|\mathcal{B}|}\sum_{n=1}^{|\mathcal{C}|}x_{i,j,m,n}(\alpha_it_{i,j,m} + \beta_ie_{i,j,m} + \gamma_i\delta_{m,n}) \leq y|S_i|/\eta_i, \forall i\in\mathcal{A},
\label{eq:30}
\end{equation}
\begin{equation}
~(5),~(6),~(7),~(8)~and~(9).
\label{eq:31}
\end{equation}
That is, the original min-max objective in OP2 is transformed into an equivalent minimization problem plus constraint (\ref{eq:30}), and this constraint denotes the sum offloading cost constraint for each MU. In spit of this, OP3 is a mixed integer non-linear programming (MINLP), thus it is NP-hard and intractable too. As that in tackling OP1, leveraging the linear relaxation, OP3 can be transformed into the following problem OP3(LR),
\begin{align*}
OP3(LR):&\min_{x_{i,j,m,n},y} y\\
s.t.&~(5),~(6),~(7)~and~(23),\\
&x_{i,j,m,n}\in[0,1],\forall i\in\mathcal{A},j\in\mathcal{S}_i,m\in\mathcal{B},n\in\mathcal{C},~y\geq 0.
\end{align*}
Since the binary variables $x_{i,j,m,n}$ of OP3 are relaxed to real numbers $x_{i,j,m,n}\in[0,1]$, then the problem OP3(LR) is linear programming and the interior points method (linear relaxation algorithm for fairness based problem, LRF) can be used to handle OP3(LR) also. However, as that for the OP1(LR), the algorithm LRF is the performance bound of our following proposed suboptimal low complexity algorithm.

\subsection{Fairness based greedy algorithm (FGA)}
First, we note that the approach of exhaustive search is still valid for OP2, but its complexity is $(|\mathcal{B}||\mathcal{C}|)^{\sum_{i\in\mathcal{A}}S_i}$, as the same as that for OP1. Thus, we focus on low complexity suboptimal algorithm and propose the fairness based greedy algorithm (FGA) for OP2. In specific, the idea of the FGA comes from the algorithm CGA and the transformed problem format OP3, and the implementation of FGA is based on an VDC too. As the CGA, the VDC collects necessary information and performs the algorithm, and finally feedbacks the results to all MUs. In order to better understand the thoughts of the FGA, we first analyze the reasons why the CGA is unfair.

From the algorithm illustration, we know that at each step, only one task offloading decision is made in the CGA. Defining the offloading cost for task $s_{i,j}$ is $u_{m,n}^{i,j}(l)$ if it is offloaded from AP $m$ to ECS $n$, and the offloading decision is made at the $l$th step. In addition, for two tasks, i.e., $s_{i_1,j}$ and $s_{i_2,j}$, their optimal offloading costs are $\hat{u}_{m_1,n_1}^{i_1,j}(l_1) = \min_{m\in\mathcal{B},n\in\mathcal{C}}u_{m,n}^{i_1,j}$ and $\hat{u}_{m_2,n_2}^{i_2,j}(l_2) = \min_{m\in\mathcal{B},n\in\mathcal{C}}u_{m,n}^{i_2,j}$, respectively, i.e., task $s_{i_1,j}$ is offloaded from AP $m_1$ to ECS $n_1$ at the $l_1$th step and task $s_{i_2,j}$ is offloaded from AP $m_2$ to ECS $n_2$ at the $l_2$th step, then we have the conclusions below.

\emph{Proposition 5}: For the algorithm CGA, if $\hat{u}_{m_1,n_1}^{i_1,j}(l_1) \leq \hat{u}_{m_2,n_2}^{i_2,j}(l_2)$, we have $l_1\leq l_2$, i.e., the offloading decision of the task $s_{i_1,j}$ precedes the task $s_{i_2,j}$.

\emph{Proposition 6}: For the algorithm CGA, defining the offloading cost for the task $s_{i,j}$ at the step $l$ is $\hat{u}^{i,j}(l)$, then $\hat{u}^{i,j}(l)$ is a non-decreasing function of $l$.

\textit{Proof:} At first, at the $l$th step, the offloading decision problem for task $s_{i,j}$ of CGA is,
\begin{equation}
OP-CGA(l): \hat{u}^{i,j}(l)=\min_{x_{i,j,m,n}} u_{m,n}^{i,j}(l)
\label{eq:32}
\end{equation}
\begin{equation}
s.t.~~x_{i,j,m,n}\in\{0,1\},m\in\hat{\mathcal{B}}^{i,j}(l),n\in\hat{\mathcal{C}}^{i,j}(l),l=1,2,...,\sum_{i\in\mathcal{A}}|\mathcal{S}_i|
\label{eq:33}
\end{equation}
Here, $\hat{\mathcal{B}}^{i,j}(l)$ and $\hat{\mathcal{C}}^{i,j}(l)$ are the accessible AP and ECS set for task $s_{i,j}$ at the step $l$, which have defined in (14) and (15), respectively. From the illustration of the algorithm CGA, it is not difficult to conclude that, both $Q_m(l), m\in\mathcal{B}$ and $R_n(l), n\in\mathcal{C}$ are non-increasing functions of $l$, i.e., with the increase of $l$, at lest one of the APs' communication resource and ECSs' computing resource are decreased. Therefore, we define the feasible solution set of $OP-CGA(l)$ as $\mathcal{F}_{CGA}(l)$ and if $l_1 < l_2$, we have $\mathcal{F}_{CGA}(l_2)\subseteq \mathcal{F}_{CGA}(l_1)$, i.e., the increase of $l$ may lead to a smaller volume of the feasible solution set, which then cloud cause the increase of the optimal value for the problem $OP-CGA(l)$ \cite{38}, i.e.,  $\hat{u}^{i,j}(l_1)\leq \hat{u}^{i,j}(l_2)$. $\Box$

Proposition 5 indicates that, whenever a task has the smaller optimal offloading cost, its offloading decision will be made ahead of other tasks. Therefore, if all the tasks of some MUs always have smaller optimal offloading cost than others, their offloading decisions for these tasks will precede others, and then from Proposition 6, this behavior results smaller offloading cost vice versa, i.e., though the greedy idea improves the offloading efficiency, it causes serious un-fairness between MUs. Therefore, a fair scheme needs break the order of the MU's offloading decision scheduling. However, this scheme should not cause much lose of the efficiency. With the above analysis and based on the target of the problem OP3, a jointly MU scheduling and task offloading decision scheme is proposed, i.e., the FGA algorithm. In which, for the MU scheduling, we design the priority function (PF) for each MU as
\begin{equation}
\chi^{i}(l) = (Y|\mathcal{S}_i|/\eta_i-\sum_{j\in\hat{\mathcal{S}}^{i}(l)}u^{i,j})/ |\bar{\mathcal{S}}^i(l)|,
\label{eq:34}
\end{equation}
where $\chi^i(l)$ is the priority function of MU $i$ in the $l$th step of the algorithm, $Y$ is a large constant, $\hat{\mathcal{S}}^i(l)$ and $\bar{\mathcal{S}}^i(l)$ are the offloaded and un-offload task set for MU $i$ before the $l$th step of the algorithm, and $u^{i,j}$ is the offloading cost of task $j$ in the set $\hat{\mathcal{S}}^i(l)$.
In practical, we can set $Y = \max_{i\in\mathcal{A},j\in\mathcal{S}_i,m\in\mathcal{B}} t_{i,j,m}+ \max_{i\in\mathcal{A},j\in\mathcal{S}_i,m\in\mathcal{B}} e_{i,j,m}+\max_{m\in\mathcal{B}, n\in\mathcal{C}} \delta_{m,n}$. For (\ref{eq:34}), $Y|\mathcal{S}_i|/\eta_i$ is from the right side of the inequation (\ref{eq:30}) which denotes the target of the offloading decision for MU $i$, $\sum_{j\in\hat{\mathcal{S}}^i(l)}u^{i,j}$ represents the accumulation offloading cost for the offloaded tasks of MU $i$, and the denominator is the number of unoffloaded tasks. The value of PF $\chi^i(l)$ means that, in order to finally attain the accumulation offloading cost $Y|\mathcal{S}_i|/\eta_i$ for the MU $i$, the average offloading cost of its un-offloaded tasks should achieve from now to the end of the algorithm. Therefore, the smaller of $\chi^i(l)$, the more strict of this constraint for the tasks of MU $i$ in $\bar{\mathcal{S}}^i(l)$. However, from the Proposition 6, we know that the offloading cost is a non-decreasing function of the step number $l$. Hence, in order to make sure that the MU's sum offloading cost constraint is satisfied with higher probability as possible, in each offloading decision period, i.e., at the $l$th step, the MU with the smallest value of $\chi^i(l)$ should be scheduled for task offloading decision, i.e., the smaller of $\chi^i(l)$, the higher priority of MU $i$. This is the philosophy of the MU scheduling.

\begin{algorithm}[h]
\caption{: Fairness based greedy algorithm (FGA)}
    \begin{algorithmic}[1]
        \STATE Initialize $l=0$, $Q_m(l)=Q_m$, $R_n(l) = R_n$, $\bar{\mathcal{A}}(l)=\mathcal{A}$,  $\bar{\mathcal{S}}^{i}(l) = \mathcal{S}_{i}$, $\hat{\mathcal{S}}^{i}(l) = \Phi$ and $Y = \max_{i\in\mathcal{A},j\in\mathcal{S}_i,m\in\mathcal{B}} t_{i,j,m}+ \max_{i\in\mathcal{A},j\in\mathcal{S}_i,m\in\mathcal{B}} e_{i,j,m}+\max_{m\in\mathcal{B}, n\in\mathcal{C}} \delta_{m,n}$;

        \STATE For $\forall i\in\bar{\mathcal{A}}(l)$, calculating $\chi^{i}(l) = (Y|\mathcal{S}_i|/\eta_i-\sum_{j\in\hat{\mathcal{S}}^{i}(l)}u^{i,j})/ |\bar{\mathcal{S}}^i(l)|$ and $i^{\star} = \arg\min_{i\in\bar{\mathcal{A}}(l)}\chi^i(l)$;

        \STATE For MU $i = i^{\star}$, we implements the following steps:
        \begin{enumerate}
            \item For $\forall j\in\bar{\mathcal{S}}^i(l)$, we calculate $\hat{\mathcal{B}}^{i,j}(l)$, $\hat{\mathcal{C}}^{i,j}(l)$ and $\Delta^{i,j}(l)$ by (\ref{eq:14}), (\ref{eq:15}) and (\ref{eq:16}), respectively;

            \item Performs \textbf{Algorithm 1} for $\forall j\in\bar{\mathcal{S}}^i(l)$, i.e., $(m_o^{i,j},n_o^{i,j},u_{o}^{i,j})=f(\hat{\mathcal{B}}^{i,j}(l),\hat{\mathcal{C}}^{i,j}(l), \Delta^{i,j}(l))$;

            \item Calculate $j_{o}^i = \arg\min_{j\in\bar{\mathcal{S}}^{i}(l)}u_{o}^{i,j}$, and let $u_{o}^i = u_{o}^{i,j_{o}^{i}}$, $m_{o}^i = m_{o}^{i,j_{o}^{i}}$ and $n_{o}^i = n_{o}^{i,j_{o}^{i}}$;
        \end{enumerate}

        \STATE Let $j^\star = j_{o}^{i^\star}$, $m^\star = m_{o}^{i^\star}$, $n^\star = n_{o}^{i^\star}$ and $u^\star =u_{o}^{i^\star}$, i.e., task $j^\star$ of MU $i^\star$ is offloaded from AP $m^\star$ to ECS $n^\star$;

        \STATE Update $R_{n^\star}(l) = R_{n^\star}(l) - r_{i^\star,j^\star}$, $Q_{m^\star}(l) = Q_{m^\star}(l) - 1$, $\bar{\mathcal{S}}^{i^\star}(l) = \bar{\mathcal{S}}^{i^\star}(l)\setminus j^\star$ and $\hat{\mathcal{S}}^{i^\star}(l) = \hat{\mathcal{S}}^{i^\star}(l)\bigcup j^\star$; if $\bar{\mathcal{S}}^{i^\star}(l)=\Phi$, we have $\bar{\mathcal{A}}(l) = \bar{\mathcal{A}}(l)\setminus i^\star$;

        \STATE If $\max_{n\in\mathcal{C}}R_n(l)\leq\min_{i\in\bar{\mathcal{A}}(l),j\in\bar{\mathcal{S}}^i(l)}r_{i,j}$, or $Q_m(l)<1,m\in\mathcal{B}$, or $\bar{\mathcal{A}}(l)=\Phi$, the algorithm is end; otherwise, $l = l +1$, goto step 2.
    \end{algorithmic}
\end{algorithm}

As aforementioned, based on the value of PF, we only determine the MU scheduling order. To complete the offloading decision, we should further make a decision about offloading task selection when the MU has multiple un-offload tasks. As the CGA, the greedy idea is adopted here again, i.e., the task with the smallest offloading cost is chosen for MU $i$. Therefore, we have the algorithm FGA, i.e., the Algorithm 4. At the step $l$, we first determine the MU scheduling order, i.e., to find the $i^\star$. It is unavoidable that, the operation of $\arg\min_{i\in\bar{\mathcal{A}}(l)}\chi^i(l)$ may feedback multiple MUs, e.g., for $l=0$, $\eta_{i_1} = \eta_{i_2}$ and $\forall i_1\neq i_2\in\mathcal{A}$, in this case, we randomly select one MU among them\footnote{In fact, in this case, the performance of the algorithm FGA can be further improved by selecting the MU with the smallest task offloading cost, however, this operation will increase the complexity of the algorithm thus it is not adopted here.}. Then we calculate the optimal offloading task for $i^\star$ based on the greedy idea, at the end of this step, we have $(i^\star,j^\star,m^\star,n^\star,u^\star)$ who characterizes the offloading task and MU, and also the corresponding offloading path and cost. At last, the system status are updated. The terminal criterion of the algorithm FGA is the same as that for the CGA. The algorithm complexity conclusion of the FGA can be summarized as below.

\emph{Proposition 7}: The algorithm complexity of the FGA is upper bounded by $O(\sum_{i\in\mathcal{A}}|\mathcal{S}_i| (|\mathcal{A}|^2+|\mathcal{S}_i|^2+(|\mathcal{B}|^2+|\mathcal{C}|^2)|\mathcal{S}_i|) + (\sum_{i\in\mathcal{A}}|\mathcal{S}_i|)^2+|\mathcal{C}|^2)$.

\textit{Proof:} Firstly, as the CGA, at each iteration, the algorithm FGA makes an offloading decision for one task. Thus, the algorithm will be end after no more than $\sum_{i\in\mathcal{A}}|\mathcal{S}_i|$ times' iteration.

Secondly, at each iteration of the algorithm FGA, the computing intensive operations are the sort operations in the step 2, step 3-2), step 3-3) and step 6\footnote{Though we need to perform the sort operation to initialize $Y$, its complexity is ignored in our analysis. Since this sort operation is not necessary and only performs one time. In fact, $Y$ takes any other positive constant is still valid for our algorithm.}. For the step 2, it is performed at the system-level thus has the complexity of $O(|\mathcal{A}|^2)$; for the step 3-2), i.e., the Algorithm 1, it is performed at task-level and for each task $j\in\bar{\mathcal{S}^i}(l)$, the sort operation has the complexity of $O(|\mathcal{C}|^2+|\mathcal{B}|^2)$; for the step 3-3), it is performed at the user-level and its complexity is at most $O(|\mathcal{S}_i|^2)$. As the same as the CGA, the step 6 only needs to perform one time over the whole process and has the complexity of $(\sum_{i\in\mathcal{A}}|\mathcal{S}_i|)^2+|\mathcal{C}|^2$. To sum up, the algorithm complexity of FGA is $O(\sum_{i\in\mathcal{A}}|\mathcal{S}_i| (|\mathcal{A}|^2+|\mathcal{S}_i|^2+(|\mathcal{B}|^2+|\mathcal{C}|^2)|\mathcal{S}_i|)+ (\sum_{i\in\mathcal{A}}|\mathcal{S}_i|)^2+|\mathcal{C}|^2)$. $\Box$

\section{Numerical results}
In this section, the performance of the proposed algorithms are evaluated through numerical simulations. As \cite{12}, in the simulation, the scenario parameters of the system are randomly generated, such as the number of MUs, the number of tasks per MU, the required CPU resource for each task, the delay and energy consumption of task offloading from MU to AP, and the ECS access-cost. In addition, the ELR and FLR algorithms are realized by the CVX tools \cite{40}. However, since the number of constraints of the ELR and FLR increase very fast when the number of variables increases, i.e., the number of MUs, the APs or ECSs, thus we only analyze their performance under small system. Without any further statements, the results shown in the following figures are an average of 1000 times independent scenarios.

\subsection{Performance bound of the proposed algorithms}

In this simulation, we study the offloading cost and fairness of the proposed algorithms, i.e., the CGA, FGA, ELR and FLR, under small system. We set that $|\mathcal{C}|=2$, $|\mathcal{B}|=3$, $|\mathcal{A}|=5$, $S_i=3,\forall i\in\mathcal{A}$, $t_{i,j,m}\in[2,6]$, $e_{i,j,m}\in[2,6]$, and $\delta_{m,n}\in[1,6]$ by varying the average computing resource requirements $r_{i,j}$ from 2 to 10, and the results are shown in Fig. 4 and Fig. 5.

\begin{figure}[]
\centering
\includegraphics[width=3.5in]{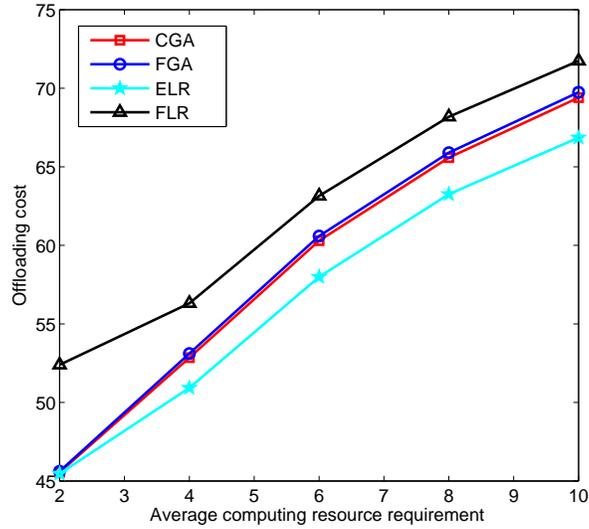}
\caption{Offloading cost comparison of the CGA, FGA, ELR and FLR.}
\label{fig.4}
\end{figure}

From Fig. 4, we note that with the increase of average computing resource requirement for each task, the offloading costs of all the proposed algorithms are increased, this comes from the fact that, the opportunity of offloading by the smallest offloading cost path decreased for each task. Moreover, as expected that, ELR has the best performance in offloading cost, then it is the algorithm CGA, while the offloading cost of the algorithm FGA is litter higher than the CGA algorithm and the FLR algorithm obtained the largest offloading cost. In addition, when the average computing resource requirement is small, e.g., $r_{i,j}\leq 2$, the offloading cost performance of both the algorithm CGA and FGA converge to the lower bound ELR. And with the increase of this average computing resource requirements, the offloading cost gap between the CGA or FGA and the ELR algorithm increased. This dues to the fact that, the larger of the average computing resource requirement for each task, the harder for multiple tasks supported by the same low offloading cost path or ECS.

\begin{figure}[]
\centering
\includegraphics[width=3.5in]{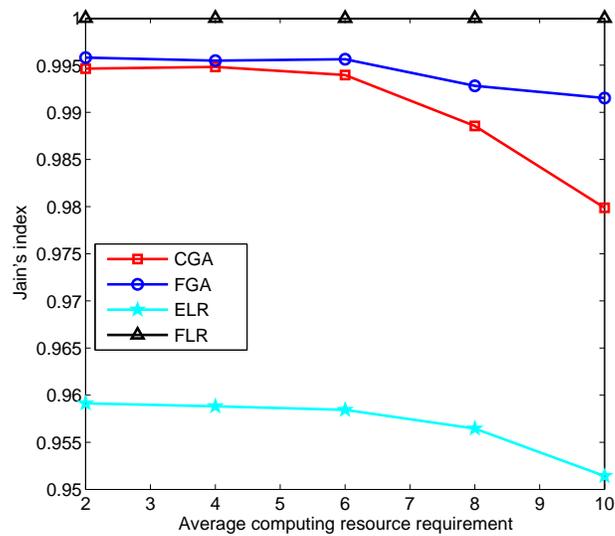}
\caption{Fairness comparison of the CGA, FGA, ELR and FLR.}
\label{fig.5}
\end{figure}

The fairness performance of the proposed algorithms is compared and shown in Fig. 5. Herein, the Jain's index fairness metric is introduced and it is defined as follows \cite{41},
$$
J(\textbf{U})=\frac{1}{n}\frac{(\sum_i U_i)^2}{\sum_iU_i^2}.
$$
For the above Jain's index fairness metric, if $J(\textbf{U})\to 1$, the algorithm is more fairness, otherwise, i.e., $J(\textbf{U})\to 0$, the algorithm is less fairness. Therefore, by observing the Fig. 5, we note that the FLR promises the best fairness performance over any other algorithms as expected, then it is our proposed algorithm FGA, and the ELR algorithm is the most unfairness algorithm. In addition, with the increase of the average computing resource requirements, the fairness performance gap between any two algorithms increased. This phenomenon can be explained as that for the Fig. 4. Moreover, from Fig. 4 and Fig. 5, we find that the ELR and the FLR are formulated the performance bound over any other algorithms in efficiency and fairness, respectively, this is consistent with the theoretical analysis.

\subsection{Comparison of the CGA, MGA and ADMA}
In Fig. 6, the offloading cost performance of the algorithm CGA and the MGA are compared under different value of $\epsilon$, i.e., $\epsilon\in[1,5]$. Herein, the simulation scenario is configured as the same as that used in Fig. 4 and Fig. 5, and we set $\eta=3$. The result in Fig.6 confirmed our analysis before that, with the varying of the value of $\epsilon$, it is possible that the MGA has less offloading cost than the CGA algorithm, i.e., $\epsilon\geq\bar{\epsilon}= 5.2$. Our further simulation results indicate that the threshold value of $\bar{\epsilon}$ depends on the value of $\eta$ and the system configurations which complies both the theory and intuition.

\begin{figure}[]
\centering
\includegraphics[width=3.5in]{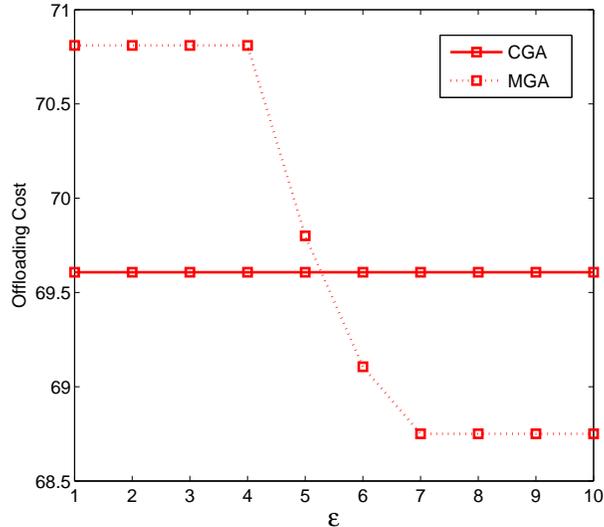}
\caption{Offloading cost comparison between CGA and MGA versus $\epsilon$.}
\label{fig.6}
\end{figure}

Then the performance comparison between the CGA and the ADMA have been done and the result is shown in Fig. 7. In the simulation, we set that $|\mathcal{C}|=4$, $|\mathcal{B}|=8$, $|\mathcal{A}|=20$, $S_i=3,\forall i\in\mathcal{A}$. The other parameters used herein are the same as that in Fig. 4 and Fig. 5. Now, the computing resource requirements for all tasks are identical, i.e., $r_{i,j}=r$, and we assumed that the mean value is varying from 2 to 10. One can observe that, the centralized algorithm CGA has better offloading cost performance over the distributed algorithm ADMA as expected. However, the difference of the offloading cost between these two algorithms is ignorable when the average computing resource requirement is small. All these results can be explained by the properties of these algorithms as mentioned earlier. Moreover, though the offloading cost performance of the ADMA is degraded, our further simulation results confirmed its performance gain over the centralized algorithm CGA in convergence rate and complexity.

\begin{figure}[]
\centering
\includegraphics[width=3.5in]{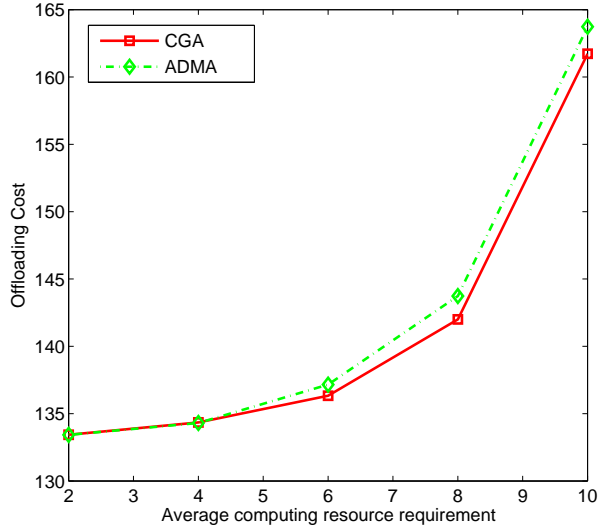}
\caption{Offloading cost comparison between CGA and ADMA.}
\label{fig.7}
\end{figure}

\subsection{More comparison between the CGA and the FGA}

Some other simulations have been done for the CGA and FGA under large system, e.g., as that in Fig. 7, we set that $|\mathcal{C}|=4$, $|\mathcal{B}|=8$, $|\mathcal{A}|=20$, and the number of tasks per MU is varying from 2 to 10. Any other simulation parameters are the same as that used in Fig. 4 and Fig. 5. We focus on the offloading cost and the Jain's index fairness metrics and the results are given in Fig. 8 and Fig. 9. Since the increasing number of tasks per MU causes the raising up of computing resource requirements in the system, the offloading cost and fairness performance of these two algorithms have similar phenomenons as that in the small system, i.e., the Fig. 4 and Fig. 5. However, it is surprising that the fairness behavior of the algorithm FGA is opposite to that shown in Fig. 5, i.e., with the increasing number of tasks per MU, the fairness performance of the FGA becomes better. Though it seems hard to understand, by retrospecting our MU scheduling PF design presented in (27), we know it is reasonable, i.e., the more tasks per MU, the scheduling degree of freedom is increased and thus better fairness performance can be promised.

\begin{figure}[]
\centering
\includegraphics[width=3.5in]{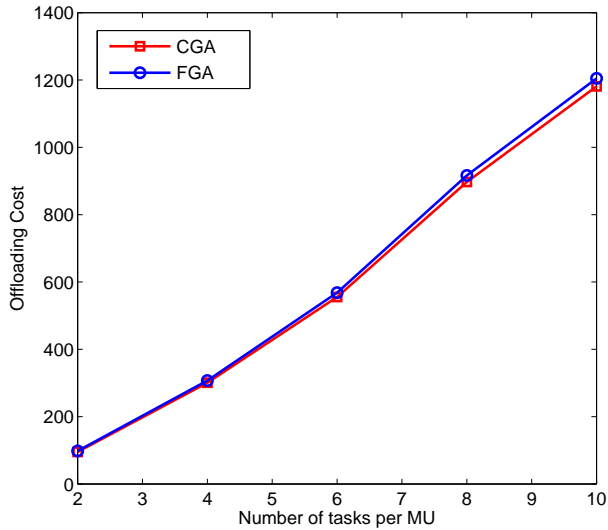}
\caption{Offloading cost comparison between CGA and FGA.}
\label{fig.8}
\end{figure}

\begin{figure}[]
\centering
\includegraphics[width=3.5in]{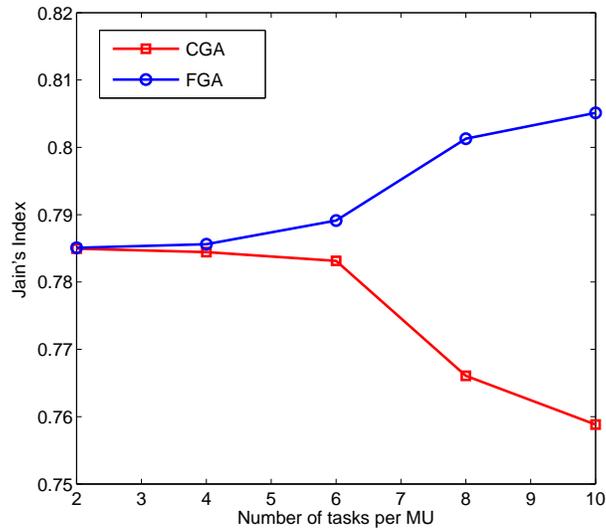}
\caption{Fairness comparison between CGA and FGA.}
\label{fig.9}
\end{figure}

At last, we analyze the offloading ratio of the algorithm CGA and FGA with limited system computing resources, i.e., the constraint (11) does not hold anymore, and the result is presented in Fig. 10. The configurations of the system are similar as that adopted in Fig. 8 and Fig. 9 except right now, the number of MUs in the system are varying from 1 to 8. Due to limited computing resources at these ECSs, only a subset of the tasks can be supported or offloaded by the system. The offloading ratio is defined as the number of successfully offloaded tasks over the total number of tasks in the system. From Fig. 10, we can note that, with the increasing number of MUs, the offloading ratio decreased for given system computing resource constraint. Surely, this is consistent with the intuition and analysis. In addition, we note that the offloading ratio gap between these two algorithm increased, which proved the efficiency of the algorithm CGA over the algorithm FGA in resource using.

\begin{figure}[]
\centering
\includegraphics[width=3.5in]{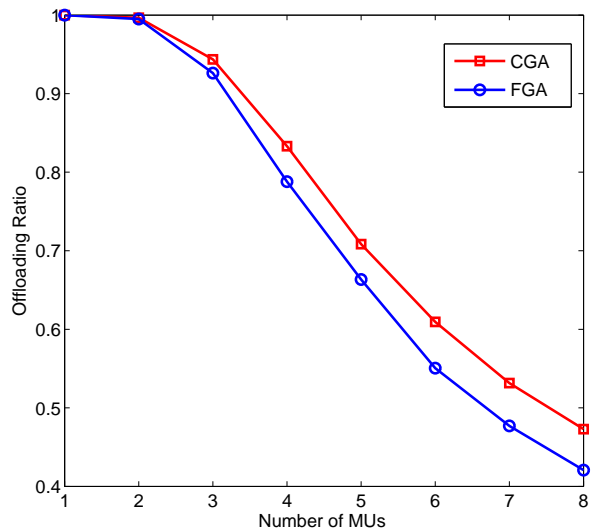}
\caption{Offloading ratio comparison between CGA and FGA.}
\label{fig.10}
\end{figure}

\section{Conclusions}
In this paper, for the imbalance edge cloud based computing offloading with multiple MUs and multiple tasks per MU, the jointly offloading decision and resource allocation problem was discussed. In particular, we proposed a new metric in counting the MU's offloading performance, i.e., the delay-energy-cost tradeoff based offloading cost. In which, the ECS access-cost was introduced to characterize the ECS access delay and the resource using payments by considering different service level agreements between APs and ECSs. Both problems of minimizing the sum offloading cost for all MUs (efficiency-based) and minimizing the maximal offloading cost per user (fairness-based) were presented. Since these problems are all NP-hard, thus several suboptimal algorithms have been proposed, i.e., the centralized algorithm CGA, MGA and the distributed algorithm ADMA for the efficiency based problem, and the algorithm FGA for the fairness based problem. Finally, our simulation results confirmed the performance of these algorithms, such as the efficiency, fairness and the complexity. Our future interests are about the incentive mechanism design for hybrid cloud system and the virtualized wireless networks.

\end{document}